\documentclass[12pt]{article}

\usepackage{dsfont}
\usepackage{epsfig}
\usepackage{amsfonts}
\usepackage{amssymb}
\usepackage{bbm}
\usepackage{cite}
\def\beq{\begin{equation}}
\def\eeq{\end{equation}}
\def\beqa{\begin{eqnarray}}
\def\eeqa{\end{eqnarray}}

\def\mZ{\mathcal{Z}}

\def\gtrsim{\mathrel{\raise.3ex\hbox{$>$\kern-.75em\lower1ex\hbox{$\sim$}}}}

\begin{document}
\pagestyle{plain}

%----------------------------------------------------------------------%
%  numbering equations with section number
%----------------------------------------------------------------------%
\makeatletter
\@addtoreset{equation}{section}
\makeatother
\renewcommand{\theequation}{\thesection.\arabic{equation}}
%----------------------------------------------------------------------%
%  title page
%----------------------------------------------------------------------%
\pagestyle{empty}
%\vspace*{1.0in}
\rightline{IFT-UAM/CSIC-08-73}
\vspace{5mm}
\begin{center}
\LARGE{\bf Non-geometric flux vacua, S-duality and algebraic geometry
\\[5mm]}
\large{
Adolfo Guarino${}^a$ and George James Weatherill${}^b$
 \\[3mm]}
\small{
${}^a$
Instituto de F\'{\i}sica Te\'orica UAM/CSIC,\\[-0.3em]
Facultad de Ciencias C-XVI, Universidad Aut\'onoma de Madrid, \\[-0.3em]
Cantoblanco, 28049 Madrid, Spain\\[2mm]

${}^b$
School of Physics and Astronomy, University of Southampton,\\[-0.3em]
Highfield, Southampton SO17 1BJ, UK\\[1cm]
}
\small{\bf Abstract} \\[3mm]
\end{center}
{\small   The four dimensional gauged supergravities descending from non-geometric string compactifications involve a wide class of flux objects which are needed to make the theory invariant under duality transformations at the effective level. Additionally, complex algebraic conditions involving these fluxes arise from Bianchi identities and tadpole cancellations in the effective theory. In this work we study a simple T and S-duality invariant gauged supergravity, that of a type IIB string compactified on a $\mathbb{T}^{6}/\mathbb{Z}_2 \times \mathbb{Z}_2$ orientifold with O3/O7-planes. We build upon the results of recent works and develop a systematic method for solving all the flux constraints based on the algebra structure underlying the fluxes. Starting with the T-duality invariant supergravity, we find that the fluxes needed to restore S-duality can be simply implemented as linear deformations of the gauge subalgebra by an element of its second cohomology class. Algebraic geometry techniques are extensively used to solve these constraints and supersymmetric vacua, centering our attention on Minkowski solutions, become systematically computable and are also provided to clarify the methods.          }

\vspace*{1.6cm}
%\leftline{\footnotesize November 2008}
\begin{flushleft}
\rule{165mm}{0.5pt}\\
$^{a}$\begin{footnotesize}e-mail: adolfo.guarino@uam.es\end{footnotesize} \\
$^{b}$\begin{footnotesize}e-mail: gjw@soton.ac.uk\end{footnotesize}
\end{flushleft}

\newpage
%----------------------------------------------------------------------%
%  Resetting of counters
%----------------------------------------------------------------------%
\setcounter{page}{1}
\pagestyle{plain}
\renewcommand{\thefootnote}{\arabic{footnote}}
\setcounter{footnote}{0}
%----------------------------------------------------------------------%
%  Paper begins
%----------------------------------------------------------------------%

\tableofcontents

\section{Introduction.}

Fluxes have played an important role in string theory research since the second string theory revolution in the mid 1990s. Orbifolds and their later extension, orientifolds, provide explicit constructions of spaces which are intimately linked to Calabi Yau manifolds but allow for specific calculation of the dynamics of the space and the fields which live within them. By constructing $\mathcal{N}=1$ orientifolds from, in the case of this paper, type IIB string compactifications, important properties and dynamics of the space can be investigated in a background which is easier to describe than that of the Calabi Yaus.
\\

Much work \cite{Lust:2005dy,Lust:2006zg} has been done on categorising the properties, such as the number of K\"{a}hler and complex structure moduli, of orbifolds constructed from six dimensional tori following the Kaluza Klein compactification of a full ten dimensional theory. In this paper we will turn our attention to one of the most studied cases, that of the $\mathbb{Z}_{2}\times \mathbb{Z}_{2}$ orbifold.
\\

A long standing problem in string theory phenomenology is the stabilization of such moduli fields. These are flat directions of the theory to all orders in perturbation theory and because of the relation between the vacuum expectation values (VEVs) of these fields with physical quantities, such as string coupling constant or internal space volumes, mechanisms for them to acquire a mass become of principal interest. One such proposed mechanism \cite{Giddings:2001yu} is based on the inclusion of background fluxes compatible with the orientifold symmetries and a lot of papers have appeared in recent years working out effective theories with a potential generated for these moduli fields as well as studying their stabilization\footnote{For reviews and references therein, see \cite{fluxes}.}. Effective theories with NS-NS, R-R and geometric $\omega$ fluxes were deeply studied in type IIA \cite{Grimm,Derendinger,vz1,DeWolfe,cfi,af}, where these fluxes were enough to stabilize all moduli in a well defined vacuum. However, this was not the case for type IIB because K\"ahler moduli do not enter in the effective potential and so remain unstabilized. To recover T-duality symmetry between type IIA and IIB string theories at the effective level, non-geometric fluxes were introduced \cite{Shelton:2005cf}.
\\

Many papers \cite{Shelton:2005cf,Shelton:2006fd,Wecht:2007wu,Micu:2007rd,Palti:2007pm} have examined the nature of these non-geometric fluxes and their vacua, as well as the mapping between type IIA and IIB scenarios \cite{Grana:2006kf,Aldazabal:2006up}. This has been done in general and also with the assumption of isotropy, where the space is a triple copy of a two dimensional toric orientifold. We will confine ourselves to the isotropic case of this orientifold but the methods discussed here readily extend to the full anisotropic orientifold. With the NS-NS background flux $\bar{H}_{3}$ turned on successive T-dualities on the six circular dimensions take the space from one having a well defined metric everywhere to one with successively more and more `pathological' descriptions, ultimately losing any notion of a locally definable metric \cite{Shelton:2005cf}. A large obstacle to investigating this and other orbifolds is that the fluxes obey non-trivial polynomial constraints through Bianchi identities of the resultant gauged Supergravity (SUGRA) \cite{Kaloper:1999yr,Hull:2005hk,Samtleben:2008pe}. Additionally, non-zero flux-induced tadpoles relate the fluxes to the localized sources living in the space. This is further complicated when we consider S-duality transformations which introduce new flux objects, as well as superpositions of fluxes of the same tensor type. This means that they each contribute to both the Bianchi and the tadpole constraints. The full set of constraints, including all non-geometric fluxes, for the deeply studied $\mathbb{Z}_{2} \times \mathbb{Z}_{2}$ orbifold are outlined in \cite{Aldazabal:2006up}.
\\

Algebraic geometry is the mathematical discipline which involves the study of polynomial systems and their solution spaces. Recent papers \cite{Gray:2006gn,Gray:2007yq,Gray:2008zs} demonstrate how previously unwieldy techniques, due to the size and complexity of the equations found in the SUGRA descriptions, of algebraic geometry can be applied to finding solutions to the aforementioned flux constraints through such programs as \textit{Singular} \cite{Singular}, thanks to the continued increase in computer speeds and memories. Useful background material for the methods used in this paper can be found in the appendix of \cite{Gray:2006gn} or the first few chapters of \cite{AGBook}. Ref. \cite{Gray:2008zs} provides a way of applying algebraic geometry to SUGRA without having to learn the specifics of \textit{Singular} or similar programs. Though the procedures outlined in \cite{Gray:2006gn,Gray:2007yq,Gray:2008zs} will be used in part, new applications of the underlying algebraic geometry methods were used in \cite{Font:2008vd} to find parametrised supersymmetric vacua in the $\mathbb{Z}_{2} \times \mathbb{Z}_{2}$ orbifold when T-duality is imposed. Unfortunately, the algorithms native to \cite{Gray:2008zs} do not immediately lend themselves to some of these methods. However, there is another interface \cite{Interface} between \textit{Mathematica} and \textit{Singular} which allows for direct access to many of \textit{Singular}'s algorithms and the methods in this paper and \cite{Font:2008vd} can be implemented using it.
\\

The present work mainly follows the line started in \cite{Font:2008vd} and extends it to include the new non-geometric flux objects induced by S-duality when both S and T dualities are considered. These algebraic geometry techniques are explained in more detail and also a new insight into the role played by these new S-duality induced fluxes is presented.
\\

The structure of this paper is the following: The starting point is a type IIB string theory compactified on a simple $\mathbb{T}^{6}/\mathbb{Z}_{2} \times \mathbb{Z}_{2}$ orientifold with O3/O7-planes. In section 2 we define our notations and conventions, which are mainly those of \cite{Font:2008vd}. Non-geometric $Q^{ab}_{c}$ and $P^{ab}_{c}$ flux objects, induced by T and S-duality transformations respectively, are then introduced. A rederivation of the T and S-duality invariant four dimensional effective theory is given, both in general and in the isotropic case. Section 3 is a short explanation of the main results concerning algebra structures, tadpole cancellation relations and methods in \cite{Font:2008vd} for the T-duality invariant effective theory. The methods used later in the paper directly derive from the methodology of \cite{Font:2008vd} and for the sake of completeness are included here. Section 4 describes the implications that S-duality transformations have for Bianchi constraints as well as for tadpole cancellation conditions from an algebraic point of view. Section 5 is devoted to clarifying the role played by the S-duality induced $P$ flux as well as a reinterpretation of Bianchi constraints involving the non-geometric fluxes in terms of integrability and cohomology conditions. We are able to solve these conditions and obtain geometric restrictions on certain modular variables induced by the non-geometric fluxes. In section 6 we solve the remaining singlet Bianchi constraint and find that two types of non-geometric flux configurations exist, labelled as type A and B configurations. Finally, in section 7, we give a family of supersymmetric $\textrm{AdS}_4$ solution without flux-induced tadpoles, followed by a systematic search of supersymmetric Minkowski solutions in terms of parametrised families of vacua. Some of them are related to previous results found in the literature.

\section{The $\mathbb{T}^{6}/\mathbb{Z}_{2}\times \mathbb{Z}_{2}$ IIB orientifold with O3/O7-planes.}

In this section we introduce the conventions adopted throughout this work and rederive the four dimensional type IIB effective field theory worked out in \cite{Aldazabal:2006up} that becomes invariant under both T and S-duality transformations.

\subsection{Conventions and 3-form fluxes.} 

We start by constructing our orbifold and defining our notation, much of which is similar or identical to that put forth in \cite{Shelton:2005cf,Shelton:2006fd,Wecht:2007wu,Font:2008vd}. We are working with a type IIB string compactification on the $\mathbb{Z}_{2} \times \mathbb{Z}_{2}$ orbifold considered in \cite{Shelton:2005cf,Aldazabal:2006up,Font:2008vd}, where the space-time topology is $\mathcal{M}_{10}=\mathcal{M}_{4} \times \mathcal{M}_{6}$, $\mathcal{M}_{6}$ our compact internal space, initially a $6$-torus, $\mathbb{T}^{6}$ and $\mathcal{M}_{4}$ the noncompact space-time. 
\\

The orbifold quotient group generators act on the tangent $1$-forms $\eta^{a}$ of $\mathcal{M}_{6}$ as 
\begin{eqnarray} 
\begin{array}{ccrccc} 
 \theta_1 & \, : \, &   ( \; \eta^{1} \; , \; \eta^{2} \; , \; \eta^{3} \; , \; \eta^{4} \; , \; \eta^{5} \; , \; \eta^{6} \; ) &\rightarrow& ( \; \eta^{1} \; , \; \eta^{2} \; , \; -\eta^{3} \; , \; -\eta^{4} \; , \; -\eta^{5} \; , \; -\eta^{6} \; ) \ , \\ 
 \theta_2 & \, : \, &   ( \; \eta^{1} \; , \; \eta^{2} \; , \; \eta^{3} \; , \; \eta^{4} \; , \; \eta^{5} \; , \; \eta^{6} \; ) &\rightarrow& ( \; -\eta^{1} \; , \; -\eta^{2} \; , \; \eta^{3} \; , \; \eta^{4} \; , \; -\eta^{5} \; , \; -\eta^{6} \; )\ .
\end{array}
\label{orbifoldgroup} 
\end{eqnarray} 

There is another order-two element $\theta_3=\theta_1\,\theta_2$. This orbifold group leads to the six dimensional torus factorisation 
\begin{eqnarray} 
\begin{array}{ccccccc} 
\mathbb{T}^{6} & = &\mathbb{T}_{1}^{2} &\times& \mathbb{T}_{2}^{2} &\times& \mathbb{T}_{3}^{2}\, . 
\end{array} 
\end{eqnarray} 

We choose our basis $\eta^{a}$ with this factorisation in mind, 
\begin{eqnarray} 
\begin{array}{cccccccccc} 
\mathbb{T}^{6} & \, &= & \, & \mathbb{T}^{2}_{1} &\times& \mathbb{T}_{2}^{2} &\times& \mathbb{T}_{3}^{2}  \ . \\ 
&  &  &  & (\eta^{1} \; , \; \eta^{2}) & & (\eta^{3} \; , \; \eta^{4}) & & (\eta^{5} \; , \; \eta^{6}) 
\end{array} 
\end{eqnarray} 

To make the space into an orientifold we quotient the space by an extra $\mathbb{Z}_2$ involution action, $\sigma$ \cite{Aldazabal:2006up,Font:2008vd},
\begin{eqnarray} 
\begin{array}{ccccc} 
\sigma \, : \, ( \; \eta^{1} \; , \; \eta^{2} \; , \; \eta^{3} \; , \; \eta^{4} \; , \; \eta^{5} \; , \; \eta^{6} \; ) &\rightarrow& ( \; -\eta^{1} \; , \; -\eta^{2} \; , \; -\eta^{3} \; , \; -\eta^{4} \; , \; -\eta^{5} \; , \; -\eta^{6} \; ) \ .
\end{array} 
\end{eqnarray}
This results in an orientifold which has three K\"{a}hler moduli and three complex structure moduli parameterizing the size and shape of the internal space.
\\

The method and ramifications in terms of supersymmetry multiplets and string actions are discussed in depth in \cite{Wecht:2007wu}. The orientifold creates O-planes which contribute to tadpole constraints. Aside from the O3, O6 and O9 planes discussed in \cite{Shelton:2005cf,Shelton:2006fd,Wecht:2007wu}, there are O4 and O7 planes discussed in \cite{Aldazabal:2006up,Font:2008vd}.
\\

Under the $\mathbb{Z}_2 \times \mathbb{Z}_2$ orbifold action, the invariant 3-forms are
\begin{eqnarray} 
\label{3formbasis}
\begin{array}{llll} 
\alpha_{0} = \eta^{135} \;&\; \alpha_{1} = \eta^{235} \;&\; \alpha_{2} = \eta^{451} \;&\; \alpha_{3} = \eta^{613} \ ,\\ 
\beta^{0} = \eta^{246} \;&\; \beta^{1} = \eta^{146} \;&\; \beta^{2} = \eta^{362} \;&\; \beta^{3} = \eta^{524}  \ ,\\  
\end{array} 
\end{eqnarray}
which are all odd under the orientifold involution $\sigma$. The invariant $2$-forms and $4$-forms basis elements come in dual pairs,
\begin{eqnarray} 
\label{24formbasis}
\begin{array}{lcllll} 
\textrm{$2$-forms} \;&\; : \;&\; \omega_{1} = \eta^{12} \;&\; \omega_{2} = \eta^{34} \;&\; \omega_{3} = \eta^{56} \ , \\ 
\textrm{$4$-forms} \;&\; : \;&\; \tilde{\omega}^{1} = \eta^{3456} \;&\; \tilde{\omega}^{2} = \eta^{1256} \;&\; \tilde{\omega}^{3} = \eta^{1234}   \ ,
\end{array} 
\end{eqnarray}
and are even under $\sigma$. Here we take our basis of forms to be that used in \cite{Font:2008vd}, with notation $\eta^{abc}=\eta^{a} \wedge \eta^{b} \wedge \eta^{c}$, etc. Denoting the volume of the internal space as $\mathcal{V}_{6}$, we fix our volume orientation and normalization by 
\begin{eqnarray}
\int_{\mathcal{M}_{6}} \eta^{123456} = \mathcal{V}_{6}\ ,
\end{eqnarray}
and so the basis satisfies
\begin{equation}
\int_{\mathcal{M}_{6}} \alpha_{0} \wedge \beta^{0} = -\mathcal{V}_{6} \quad , \quad \int_{\mathcal{M}_{6}} \alpha_{I} \wedge \beta^{J} = \int_{\mathcal{M}_{6}} \omega_{I} \wedge \tilde{\omega}^{J} = \mathcal{V}_{6} \; \delta_{I}^{J} \,\,\,\,,\,\,\,\,I,J=1,2,3.
\end{equation}

With these properties the holomorphic $3$-form of the compact internal space, $\Omega$, has a convenient expansion in terms of the 3-form basis (\ref{3formbasis}),
\begin{eqnarray}
\Omega = (\eta^{1}+\tau_{1}\eta^{2})\wedge(\eta^{3}+\tau_{2}\eta^{4})\wedge(\eta^{5}+\tau_{3}\eta^{6}) = \alpha_{0}+\tau_{K}\alpha_{K} + \beta^{K}\frac{\tau_{1}\tau_{2}\tau_{3}}{\tau_{K}} + \beta^{0}\tau_{1}\tau_{2}\tau_{3} \, .
\end{eqnarray}

The NS-NS $H_{3}$ and the R-R $F_{3}$ fields are odd under the orientifold action $\sigma$. Then, consistent background fluxes can be expanded in terms of (\ref{3formbasis}) as
\beqa
\bar H_3 & = & b_{3} \,\alpha_{0} + b_2^{(I)} \,\alpha_{I}  + b_{1}^{(I)} \,\beta^{I} + b_{0} \,\beta^{0} \ , 
\label{H3expan} \\[2mm]
\bar F_3 & = & a_{3} \,\alpha_{0} + a_{2}^{(I)}  \,\alpha_{I}  + 
a_{1}^{(I)} \,\beta^{I} + a_{0} \,\beta^{0} \ ,
\label{F3expan} 
\eeqa
with $I=1,2,3$. All flux coefficients are integers because the integrals of  $\bar{H}_{3}$ and $\bar{F}_{3}$ over 3-cycles are quantized. To avoid subtleties with exotic orientifold planes
we take all fluxes to be even \cite{frey, kst}.

\subsection{Moduli, dual fluxes and flux-induced superpotential.}

We define our moduli fields to be $S = C_{0}+ie^{-\phi}$, the 4d complex axiodilaton, where $C_{0}$ is the R-R $0$-form and $\phi$ the 10d dilaton, $U_{I} = \tau_{I}$ our complex structure moduli and then $T_{I}$, our K\"{a}hler moduli, via the expansion of the complexified K\"{a}hler $4$-form $\mathcal{J} = -\sum T_{I}\tilde{w}^{I}$. The K\"ahler potential, to tree level, is therefore given as
\begin{eqnarray}
K &= & -\sum_{I=1}^{3}\ln \Big( -i(T_{I}-\bar{T}_{I}) \Big) -\ln \Big( -i(S-\bar{S}) \Big) -\sum_{I=1}^{3}\ln \Big( -i(U_{I}-\bar{U}_{I}) \Big) \, .
\end{eqnarray}

Therefore, working with only the two $3$-form fluxes, the NS-NS $\bar{H}_{3}$ and the R-R $\bar{F}_{3}$, we have the standard GVW superpotential,
\begin{eqnarray}
W = \int_{\mathcal{M}_{6}} (\bar{F}_{3}-S\, \bar{H}_{3}) \wedge \Omega \, .
\end{eqnarray}

Due to the lack of $T_{I}$ moduli in this superpotential expression, it would not be possible to obtain stable vacua with all moduli stabilized without the inclusion of non-perturbative effects such as gaugino condensation. However, considerable discussion \cite{Shelton:2005cf,Shelton:2006fd,Wecht:2007wu,Aldazabal:2006up} has been done on the effect of T-duality on such orientifolds and the fluxes within them. Upon considering the algebra of the generators of gauge choices, $X^{a}$, on the NS-NS $B$ field and diffeomorphisms, $Z_{b}$, on the metric of the space under multiple T-dualities, it has been found that the orientifold possesses a twelve dimensional algebra of these generators in terms of NS-NS $\bar{H}_{3}$, geometric $\omega$ and generalized non-geometric $Q$ and $R$ fluxes. The non-geometric $Q$ flux gives rise to a background that is locally but not globally geometric while the $R$ flux yields a background that is not even locally geometric. 
\\

From now on, we will always restrict our considerations to the IIB orientifold with O3/O7-planes which excludes the geometric $\omega$ and the non-geometric $R$ fluxes and reduces the algebra to
\beqa
\label{XZAlgebra} 
\big[ Z_{a}, Z_{b} \big] &=& \bar{H}_{abc}X^{c}  \nonumber \\ 
\big[ Z_{a}, X^{b} \big] &=& Q^{bc}_{a}Z_{c}  \\ 
\big[ X^{a}, X^{b} \big] &=& Q^{ab}_{c}X^{c}  \nonumber
\eeqa
involving the NS-NS $\bar{H}_3$ and the non-geometric $Q$ background fluxes.
\\

We are able to include this new non-geometric $Q$ flux in the superpotential by contracting it with $\mathcal{J}$,
\begin{eqnarray}
(Q \cdot \mathcal{J})_{abc} = \frac{1}{2}Q_{[a}^{de}(\mathcal{J})_{bc]de} \quad \Rightarrow \quad \int_{\mathcal{M}_{6}} (Q \cdot \mathcal{J}) \wedge \Omega \subset W \, .
\end{eqnarray}

As a 3-form, $Q \cdot \mathcal{J}$ can be expanded in the basis (\ref{3formbasis}),
\beq
\label{QJexpan}
Q \cdot \mathcal{J}=T_{K} \left( c_{3}^{(K)} \,\alpha_{0} - \mathcal{C}_{2}^{(I K)} \,\alpha_{I}  - \mathcal{C}_{1}^{(I K)} \,\beta^{I} 
+ c_{0}^{(K)} \,\beta^{0}  \right) \ ,
\eeq
where $\mathcal{C}_1$ and $\mathcal{C}_2$ are the non-geometric $Q$ flux matrices
\beq
\mathcal{C}_{1}=\left(
\begin{array}{rrr}
-\tilde{c}_{1}^{\,(1)} & \check{c}_{1}^{\,(3)}   & \hat{c}_{1}^{\,(2)}    \\
 \hat{c}_{1}^{\,(3)}   & -\tilde{c}_{1}^{\,(2)}  & \check{c}_{1}^{\,(1)}  \\
 \check{c}_{1}^{\,(2)} & \hat{c}_{1}^{\,(1)}     & -\tilde{c}_{1}^{\,(3)} \\
\end{array}
\right)
\qquad ,\qquad
\mathcal{C}_{2}=\left(
\begin{array}{rrr}
-\tilde{c}_{2}^{\,(1)} & \check{c}_{2}^{\,(3)}   & \hat{c}_{2}^{\,(2)}    \\
 \hat{c}_{2}^{\,(3)}   & -\tilde{c}_{2}^{\,(2)}  & \check{c}_{2}^{\,(1)}  \\
 \check{c}_{2}^{\,(2)} & \hat{c}_{2}^{\,(1)}     & -\tilde{c}_{2}^{\,(3)} \\
\end{array}
\right) \ .
\label{c1c2mat}
\eeq

This T-duality invariant 4d effective theory involving the $\bar{H}_{3}$, $\bar{F}_{3}$ and $Q$ fluxes is described by the superpotential,
\begin{eqnarray} 
W = \int_{\mathcal{M}_{6}} (\bar{F}_{3} - S \, \bar{H}_{3} + Q \cdot \mathcal{J}) \wedge \Omega  \, .
\end{eqnarray}

However, we wish to consider S-duality on top of T-duality and in seeing how the known fluxes of the superpotential behave under S-duality, we can infer the existence of new fluxes and new constraints. The dynamics of the moduli can be computed from the bosonic scalar potential, $V_{F}$, when $D$ terms are neglected. $V_{F}$ is a functional of the K\"{a}hler function $G$, which is itself a functional of the superpotential and the K\"{a}hler potential,
\begin{eqnarray}
V_{F} = e^{G}G^{a\bar{b}}\partial_{a}G \overline{\partial_{b}G} = e^{K}\left( K^{a\bar{b}}D_{a}W \overline{D_{b}W} - 3|W|^{2} \right) \quad , \quad G \equiv K+\ln |W|^{2}\ , 
\label{VPotential}
\end{eqnarray} 
where $G^{a\overline{b}}$ is the inverse of $\partial_{a}\bar{\partial_{b}}G$ and likewise for $K^{a\overline{b}}$.
\\

We wish to impose a symmetry, that of S-duality, where the transformation has a non linear action $ S \to \frac{k \, S \,+\, l}{m \, S \,+\, n}$, defined by an element $\Theta_{S}=\pmatrix{k & l \cr m & n} \in SL(2,\mathbb{Z})_{S}$. In order for $G$ to be invariant under this transformation, the superpotential must transform in a particular way,
\begin{eqnarray}
W(S) \to W\left(\frac{kS + l}{mS + n}\right) = \frac{1}{mS+n}W(S) \, .
\end{eqnarray} 

Therefore the fluxes must themselves transform in such a way as to satisfy this and they must transform in multiplets. Therefore, having non-trivial $\bar{H}_{3}$ or $\bar{F}_{3}$ flux means allowing for both $3$-form fluxes being non-zero following such a transformation in $S$,
\begin{eqnarray} 
\bar{F}_{3}-S\,\bar{H}_{3} \to \bar{F}'_{3}-\left(\frac{kS + l}{mS + n}\right)\bar{H}'_{3} = \frac{1}{mS+n}\Big( (n \bar{F}'_{3}-l \bar{H}'_{3})-S (k \bar{H}'_{3}-m \bar{F}'_{3}) \Big) \, .
\end{eqnarray} 

Solving for $\bar{F}'_{3}$ and $\bar{H}'_{3}$ in terms of $\bar{F}_{3}$ and $\bar{H}_{3}$ we have that the S-duality action on the $3$-form fluxes is 
\begin{eqnarray} 
\label{FHtransf}
\pmatrix{\bar{F}'_{3} \cr \bar{H}'_{3}} = \Theta_{S}\pmatrix{\bar{F}_{3} \cr \bar{H}_{3}} = \pmatrix{ k\bar{F}_{3}+l\bar{H}_{3} \cr m\bar{F}_{3}+n\bar{H}_{3} } \, .
\end{eqnarray} 

Similarly, $Q$ needs to be partnered with another flux of the same tensor type and we are forced to turn on another non-geometric flux, $P$, which is multiplied by the axiodilaton in order to give the same doublet mixing, 
\begin{eqnarray}
\label{QPtransf} 
\pmatrix{Q' \cr P'} =\Theta_{S}\pmatrix{Q \cr P} = \pmatrix{ kQ+lP \cr mQ+nP} \, . 
\end{eqnarray} 

With the inclusion of this additional non-geometric $P$ flux, we are lead to a both T and S-duality invariant 4d effective theory. It involves the $\bar{H}_{3}$, $\bar{F}_{3}$, $Q$ and $P$ fluxes and is described by the superpotential,
\begin{eqnarray} 
\label{superpotential}
W  = \int_{\mathcal{M}_{6}} \Big(\bar{F}_{3} - S \, \bar{H}_{3} + (Q-S \, P) \cdot \mathcal{J}\Big) \wedge \Omega \, .
\end{eqnarray} 

Further terms occur in the superpotential if $\omega, R \not= 0$ \cite{Aldazabal:2006up}, such as in IIB with O9-planes and they must come in similar doublets. Here we will be working with the IIB orientifold with O3/O7-planes that excludes a background for these fluxes.
\\

As with $Q\cdot \mathcal{J}$, $P \cdot \mathcal{J}$ can be expanded in the $3$-form basis (\ref{3formbasis}),
\beq
\label{PJexpan}
P \cdot \mathcal{J}=T_{K} \left( d_{3}^{(K)} \,\alpha_{0} - \mathcal{D}_{2}^{(I K)} \,\alpha_{I}  - \mathcal{D}_{1}^{(I K)} \,\beta^{I} 
+ d_{0}^{(K)} \,\beta^{0}  \right) \ ,
\eeq
where $\mathcal{D}_1$ and $\mathcal{D}_2$ are the new non-geometric $P$ flux matrices,
\beq
\mathcal{D}_{1}=\left(
\begin{array}{rrr}
-\tilde{d}_{1}^{\,(1)} & \check{d}_{1}^{\,(3)}   & \hat{d}_{1}^{\,(2)}    \\
 \hat{d}_{1}^{\,(3)}   & -\tilde{d}_{1}^{\,(2)}  & \check{d}_{1}^{\,(1)}  \\
 \check{d}_{1}^{\,(2)} & \hat{d}_{1}^{\,(1)}     & -\tilde{d}_{1}^{\,(3)} \\
\end{array}
\right)
\qquad ,\qquad
\mathcal{D}_{2}=\left(
\begin{array}{rrr}
-\tilde{d}_{2}^{\,(1)} & \check{d}_{2}^{\,(3)}   & \hat{d}_{2}^{\,(2)}    \\
 \hat{d}_{2}^{\,(3)}   & -\tilde{d}_{2}^{\,(2)}  & \check{d}_{2}^{\,(1)}  \\
 \check{d}_{2}^{\,(2)} & \hat{d}_{2}^{\,(1)}     & -\tilde{d}_{2}^{\,(3)} \\
\end{array}
\right) \ .
\label{d1d2mat}
\eeq

The locations of these flux entries within the non-geometric fluxes $Q$ and $P$ are shown in table \ref{tableNonGeometric}. Their S-duality doublet partners in $P$ are found by the exchanges $Q \leftrightarrow P$ and $c \leftrightarrow d$. In line with \cite{Font:2008vd} we will use Greek indices $\alpha,\beta,\gamma$ for horizontal $\,``-"$ $x$-like directions $(\eta^{1},\eta^{3},\eta^{5})$ and Latin indices $i,j,k$ for vertical $\,``|"$ $y$-like directions $(\eta^{2},\eta^{4},\eta^{6})$ in the 2-tori and we choose our $a_{i}$ , $\ldots$ , $d_{j}$ appear in locations with positive index permutations. 
\\

The superpotential (\ref{superpotential}) depends on the seven untwisted closed string moduli and takes the form
\beq
\label{Wfluxes}
W=P_{1}(U) + P_{2}(U)\,S + \sum_{K=1}^{3} P_{3}^{\,(K)}(U)\,T_{K} + S \, \sum_{K=1}^{3} \, P_{4}^{\,(K)}(U)\,T_{K}\, ,
\label{fullW}
\eeq
involving cubic polynomials in the complex structure moduli given by
\beqa
P_{1}(U) & = & a_{0} -\sum_{K=1}^{3} a_{1}^{\,(K)}\,U_{K} + 
\sum_{K=1}^{3} a_{2}^{\,(K)} \frac{U_{1}U_{2}U_{3}}{U_{K}} - a_{3} U_{1}U_{2}U_{3}   \ ,
\label{p1gen} \\[2mm]
P_{2}(U) & = & -b_{0} +\sum_{K=1}^{3} b_{1}^{\,(K)}\,U_{K} - 
\sum_{K=1}^{3} b_{2}^{\,(K)} \frac{U_{1}U_{2}U_{3}}{U_{K}} + b_{3} U_{1}U_{2}U_{3}  \ ,  
\label{p2gen} \\[2mm]
P_{3}^{\,(K)}(U) & = & c_{0}^{\,(K)} +\sum_{L=1}^{3} \mathcal{C}_{1}^{\,(L K)}\,U_{L} - 
\sum_{L=1}^{3} \mathcal{C}_{2}^{\,(L K)} \frac{U_{1}U_{2}U_{3}}{U_{L}} -c_{3}^{\,(K)} U_{1}U_{2}U_{3}  \ ,
\label{p3gen}\\[2mm]
P_{4}^{\,(K)}(U) & = & -d_{0}^{\,(K)} -\sum_{L=1}^{3} \mathcal{D}_{1}^{\,(L K)}\,U_{L} + 
\sum_{L=1}^{3} \mathcal{D}_{2}^{\,(L K)} \frac{U_{1}U_{2}U_{3}}{U_{L}} +d_{3}^{\,(K)} U_{1}U_{2}U_{3}  \ .
\label{p4gen}
\eeqa
We have defined $P_{2}(U)$ and $P_{4}(U)$ so that the coefficient of $S$ in (\ref{Wfluxes}) is $+1$, rather than the $-1$ in (\ref{superpotential}).
\begin{table}[htb]
\begin{center}\begin{tabular}{|c|c|c|}
\hline
Type & Components & Fluxes \\
\hline
\hline
$Q_{-}^{--} \equiv Q_{\alpha}^{\beta \gamma}$ & $ Q_{1}^{35}\,,\,Q_{3}^{51}\,,\,Q_{5}^{13}$ & 
$\tilde{c}_{1}^{\,(1)}\,,\,\tilde{c}_{1}^{\,(2)}\,,\,\tilde{c}_{1}^{\,(3)}$ \\
\hline
\hline
$Q_{|}^{|-} \equiv Q_{k}^{i \beta} $ & $ Q_{4}^{61}\,,\,Q_{6}^{23}\,,\,Q_{2}^{45}$ & 
$ \hat{c}_{1}^{\,(1)}\,,\,\hat{c}_{1}^{\,(2)}\,,\,\hat{c}_{1}^{\,(3)}$ \\
\hline
\hline
$Q_{|}^{-|} \equiv Q_{k}^{\alpha j}$ & $ Q_{6}^{14}\,,\,Q_{2}^{36}\,,\,Q_{4}^{52}$ & 
$ \check{c}_{1}^{\,(1)}\,,\,\check{c}_{1}^{\,(2)}\,,\,\check{c}_{1}^{\,(3)}$ \\
\hline
\hline
$Q_{|}^{--} \equiv Q_{k}^{\alpha\beta}$ & $Q_{2}^{35}\,,\,Q_{4}^{51}\,,\,Q_{6}^{13}$ &  
$ c_{0}^{\,(1)}\,,\,c_{0}^{\,(2)}\,,\,c_{0}^{\,(3)}$\\
\hline
\hline
$Q_{-}^{||} \equiv Q_{\gamma}^{i j}$ & $ Q_{1}^{46}\,,\,Q_{3}^{62}\,,\,Q_{5}^{24}$ & 
$ c_{3}^{\,(1)}\,,\,c_{3}^{\,(2)}\,,\,c_{3}^{\,(3)}$ \\
\hline
\hline
$Q_{-}^{|-} \equiv Q_{\gamma}^{i \beta}$ & $Q_{5}^{23}\,,\,Q_{1}^{45}\,,\,Q_{3}^{61}$ & 
$\check{c}_{2}^{\,(1)}\,,\,\check{c}_{2}^{\,(2)}\,,\,\check{c}_{2}^{\,(3)}$ \\
\hline
\hline
$Q_{-}^{-|} \equiv Q_{\beta}^{\gamma i}$ & $ Q_{3}^{52}\,,\,Q_{5}^{14}\,,\,Q_{1}^{36}$ & 
$\hat{c}_{2}^{\,(1)}\,,\,\hat{c}_{2}^{\,(2)}\,,\,\hat{c}_{2}^{\,(3)}$ \\
\hline
\hline
$Q_{|}^{||} \equiv Q_{k}^{i j}$ & $Q_{2}^{46}\,,\,Q_{4}^{62}\,,\,Q_{6}^{24}$ & 
$\tilde{c}_{2}^{\,(1)}\,,\,\tilde{c}_{2}^{\,(2)}\,,\,\tilde{c}_{2}^{\,(3)}$ \\
\hline
\end{tabular}\end{center}
\caption{Non-geometric $Q$-flux. $P$ flux is defined by replacing $c$ with $d$.}
\label{tableNonGeometric}
\end{table}

\subsection{The isotropic ansatz.} 

Looking for algebra structures behind non-geometric fluxes in the general case is a difficult task, beyond the scope of this work. Therefore, for simplicity, we are considering the isotropic orbifold, where the three 2-tori are identical, giving $\mathbb{T}^{6} = (\mathbb{T}^{2})^{3}$. This additional symmetry reduces the number of independent flux entries. From now on, we will restrict ourselves to the isotropic flux configurations shown in tables \ref{tableIsoNSRR}, \ref{tableIsoNon-GeometricQ} and \ref{tableIsoNon-GeometricP}, all compatible with the isotropic moduli vacua ansatz, $U_{i}\to U$ and $T_{i} \to T$.
\\

\begin{table}[htb]
\begin{center}\begin{tabular}{|c|c|c|c||c|c|c|c|}
\hline
$\bar{F}_{---}$ & $\bar{F}_{|--}$ & $\bar{F}_{-||}$ & $\bar{F}_{|||}$ & 
$\bar{H}_{---}$ & $\bar{H}_{|--}$ & $\bar{H}_{-||}$ & $\bar{H}_{|||}$\\
\hline
$a_{3}$ & $a_{2}$ & $a_{1}$ & $a_{0}$ & $b_{3}$ & $b_{2}$ & $b_{1}$ & $b_{0}$ \\
\hline
\end{tabular}\end{center}
\caption{R-R and NS-NS isotropic fluxes. }
\label{tableIsoNSRR}
\end{table}

\begin{table}[htb]
\begin{center}\begin{tabular}{|c|c|c|c|c|c|c|c|}
\hline
$Q_{-}^{--}$ & $Q_{|}^{|-}$ & $Q_{|}^{-|}$ & $Q_{|}^{--}$ & $Q_{-}^{||}$ & $Q_{-}^{|-}$ & $Q_{-}^{-|}$ & $Q_{|}^{||}$ \\
\hline
$\tilde{c}_{1}$ & $ \hat{c}_{1}$ & $ \check{c}_{1}$ & $ c_{0}$ & $c_{3}$ & $\check{c}_{2}$ & $\hat{c}_{2}$ & $\tilde{c}_{2}$\\
\hline
\end{tabular}\end{center}
\caption{Non-geometric $Q$ isotropic flux.}
\label{tableIsoNon-GeometricQ}
\end{table}

\begin{table}[htb]
\begin{center}\begin{tabular}{|c|c|c|c|c|c|c|c|}
\hline
$P_{-}^{--}$ & $P_{|}^{|-}$ & $P_{|}^{-|}$ & $P_{|}^{--}$ & $P_{-}^{||}$ & $P_{-}^{|-}$ & $P_{-}^{-|}$ & $P_{|}^{||}$\\
\hline
$\tilde{d}_{1}$ & $ \hat{d}_{1}$ & $ \check{d}_{1}$ & $ d_{0}$ & $d_{3}$ & $\check{d}_{2}$ & $\hat{d}_{2}$ & $\tilde{d}_{2}$\\
\hline
\end{tabular}\end{center}
\caption{Non-geometric $P$ isotropic flux.}
\label{tableIsoNon-GeometricP}
\end{table}

A further reduction in the number of flux entries is discussed in \cite{Shelton:2006fd,Font:2008vd}. We are considering real integer flux entries and in order to have $\tilde{c}_{i},\tilde{d}_{i} \in \mathbb{R}$ for $i = 1,2$, we have the constraints 
\begin{eqnarray} 
\hat{c}_{i} = \check{c}_{i} \equiv c_{i} \hspace{5mm},\hspace{5mm} \hat{d}_{i} = \check{d}_{i} \equiv d_{i} \ . 
\end{eqnarray} 

After this reduction in both the number of independent flux entries and moduli fields the superpotential (\ref{Wfluxes}) becomes
\beqa
W = P_{1}(U)+ S \, P_{2}(U)+ T \, P_{3}(U) + S \,T \, P_{4}(U) \ ,
\label{Wiso}
\eeqa
where the polynomials have simplified to
\begin{eqnarray}
\label{Polyiso}
P_{1}(U)& = & a_{0}-3 \, a_{1}U + 3\, a_{2}U^{2} - a_{3}U^{3}  \ ,\\
P_{2}(U)& = & -b_{0}+3 \, b_{1}U - 3 \, b_{2}U^{2} + b_{3}U^{3}  \ ,\\
P_{3}(U)& = & 3\,\left( \, c_{0}+(2 \, c_{1}-\tilde{c}_{1})U - (2 \, c_{2}-\tilde{c}_{2})U^{2} - c_{3}U^{3} \, \right)   \ ,\\
P_{4}(U)& = & 3\,\left( \, -d_{0}-(2 \, d_{1}-\tilde{d}_{1})U + (2 \, d_{2}-\tilde{d}_{2})U^{2} + d_{3}U^{3} \, \right) \ .
\end{eqnarray}
For convenience, the factors of $3$ arise from the summation over $T^{K}$ in (\ref{Wfluxes}) have been absorbed into the polynomials. Under the isotropic ansatz, the K\"ahler potential reduces to
\begin{eqnarray}
K &= & -3\,\ln \Big( -i(T-\bar{T}) \Big) -\ln \Big( -i(S-\bar{S}) \Big) -3\,\ln \Big( -i(U-\bar{U}) \Big) \, .
\label{Kiso}
\end{eqnarray}

\section{Overview of the T-duality invariant 4d effective theory.}

Without considering S-duality, we restrict the system to having $P=0$. This case has been deeply studied in \cite{Font:2008vd} and here we present the main results as well as the methods used. These methods will become crucial to exploring the algebraic structure once the $P$ flux has been included.

\subsection{Fluxes and Lie algebras: The results.} 

The coefficients appearing in the polynomials are subject to certain Lie algebra constraints because the fluxes are structure constants in the algebra (\ref{XZAlgebra}). Recent work \cite{Font:2008vd} has provided a method to address these constraints in a general and algorithmic way and which readily extends to other orbifolds. 
\\

Working in type IIB with O3/O7-planes, we first address the constraint equations generated by the remaining non-zero fluxes, as derived in \cite{Wecht:2007wu}
\begin{eqnarray}
\label{BianchiT}
Q^{[ab}_{e}Q^{c]e}_{d} \equiv QQ = 0 \hspace{7mm},\hspace{7mm} Q^{ed}_{[a}\bar{H}_{bc]d} \equiv Q\bar{H}_{3} =  0 \, .
\end{eqnarray} 
These constraints can be viewed in two manners. Firstly, they can be viewed as generated by the commutation relations in (\ref{XZAlgebra}) needing to satisfy Jacobi identities for Lie algebras. For instance if the three generators in the Jacobi identity are all $X$ generators then the equation becomes the first constraint above. This is of particular interest to us because the $X^{a}$ form a six dimensional subalgebra. Secondly, they can be viewed as Bianchi constraints which arise from requiring nilpotency, $D^{2}=0$, on the operator $D= d+\bar{H}_{3}\wedge \,+\,Q \cdot$ on the torus, where the fluxes act as contributions to torsion, as discussed in \cite{Shelton:2006fd}. This second method is used in \cite{Ihl:2007ah} to solve such constraints on a $\mathbb{Z}_{4}$ orbifold.
\\

First we focus on $Q^{ab}_{c}$, which has the additional properties $Q^{ab}_{b}=0$ and $Q^{ab}_{c} = -Q^{ba}_{c}$ and is playing the role of a structure constant in a six dimensional $X^{a}$ gauge subalgebra of (\ref{XZAlgebra}). In terms of the $Q$ flux entries, $QQ=0$ becomes
\beqa
\label{BianchiC}
c_0 \left(c_2-\tilde{c}_2\right)+ c_1\,(c_1-\tilde{c}_1) &=&0 \ , \nonumber \\
c_2\,(c_2-\tilde{c}_2)+c_3 \left(c_1-\tilde{c}_1\right)  &=&0  \ , \\
c_0 c_3-c_1 c_2 &=&0  \ . \nonumber 
\eeqa

Due to the isotropic orientifold symmetries, the internal space tangent form basis $\eta^{a}$ can be split into two $3$ dimensional systems, $\eta^{a} \to (\xi^{I},\tilde{\xi}^{I})$, which are invariant under the permutation $\xi^{1} \to \xi^{2} \to \xi^{3} \to \xi^{1}$, and similarly for $\tilde{\xi}^{I}$, which is required for isotropy. Also, the Cartan-Killing metric built from the $Q$ flux by
\beq
\label{Killing-Cartan}
\mathcal{H}^{ab}= Q^{ad}_{c} Q^{bc}_{d} \ 
\eeq
has a $3+3$ block diagonal structure. There are only five isotropic non-trivial Lie algebras with such properties; $\mathfrak{so(4)} \sim \mathfrak{su(2)^{2}}$, $\mathfrak{so(3,1)}$, $\mathfrak{su(2)} + \mathfrak{u(1)^{3}}$, $\mathfrak{iso(3)}$ and $\mathfrak{nil}$\footnote{where $\mathfrak{nil} \equiv \,\,n\,$3.5 in \cite{Grana:2006kf}. A complete classification of six dimensional real nilpotent Lie algebras is given in \cite{Salamon}.}. We do not consider the abelian $\mathfrak{u(1)^{6}}$ since it is equivalent to a trivial $Q=0$ background flux. All these algebras are quasi-classical Lie algebras, ie. they have an invariant non-degenerate metric built from their quadratic Casimir operator \cite{CampoamorStursberg}. In the redefined $1$-form $(\xi^{I},\tilde{\xi}^{I})$ basis, these algebras have the canonical form shown in table \ref{canonicQalgebras}, where an antisymmetric structure $\epsilon_{IJK}$ is always understood.
\\

\begin{table}[htb]
\small{
\renewcommand{\arraystretch}{1.15}
\begin{center}\begin{tabular}{|c|c|c|}
\hline
Algebra                                &  $d \xi^I$                                           &  $d \tilde{\xi}^I$ \\
\hline
\hline
$\mathfrak{so(4)} \sim \mathfrak{su(2)^{2}}$  &    $ \,\xi^J \wedge \xi^K$             &     $ \,\tilde{\xi}^J \wedge \tilde{\xi}^K$ \\
\hline
$\mathfrak{so(3,1)}$ & $ \,\xi^J \wedge \xi^K - \tilde{\xi}^J \wedge \tilde{\xi}^K $ & $ \,\xi^J \wedge \tilde{\xi}^K$ \\
\hline
$\mathfrak{su(2)} + \mathfrak{u(1)^3}$ &    $ \,\xi^J \wedge \xi^K$             &     0 \\
\hline
$\mathfrak{iso(3)}$  & $ \,\xi^J \wedge \xi^K$  &  $ \, \xi^J \wedge \xi^K + \xi^J \wedge \tilde{\xi}^K $ \\
\hline
$\mathfrak{nil}$                       &    0             &     $ \,\xi^J \wedge \xi^K$ \\
\hline
\end{tabular}\end{center}
\caption{Canonic non-geometric $Q$ algebras. }
\label{canonicQalgebras}
}
\end{table}

A complete study of the vacua relating to each algebra was performed in \cite{Font:2008vd}. These were also related to the localized sources that have to be present in the theory.
\\

Once the $Q$ flux is chosen to be the transformed structure constant of one of the above algebras, the NS-NS flux is easily calculated from the system $Q\bar{H}_{3}=0$, which is linear on the NS-NS flux entries $b_i$,
\beq
\label{BianchiB}
\begin{array}{lll}
-c_2 \, b_0 + (c_1-\tilde{c}_1) \, b_1 + c_0 \, b_2    &=&0 \ ,  \\
-c_2 \, b_1 + (c_1-\tilde{c}_1) \, b_2 + c_0 \, b_3    &=&0 \ ,  \\
-c_3 \, b_0 - (c_2-\tilde{c}_2) \, b_1 + c_1 \, b_2    &=&0   \ , \\
-c_3 \, b_1 - (c_2-\tilde{c}_2) \, b_2 + c_1 \, b_3    &=&0 \ .  
\end{array}
\eeq

The problem is then reduced to computing the 2 dimensional kernel of this linear system.

\subsection{Fluxes and Lie algebras: The methods.} 

The method of finding a parametrised solution to (\ref{BianchiT}) consists of selecting one of the Lie algebras, $\mathfrak{g}$, in table \ref{canonicQalgebras} and reading its canonical structure constants, $g_{I}^{JK}$, from there. Then, by performing a change of basis on the algebra\footnote{Instead of changing the 1-form basis $M^{-1}: (\eta^{a}) \to (\xi^{I},\tilde{\xi}^{I})$, we will move to its dual, the $X$ generators basis, with the transformation $M: (X^{a}) \to (E^{I},\tilde{E}^{I})$.},
\beqa
\label{equivalgebras}
Q= M^{-1} \, M^{-1} \, g \,\, M \ ,
\eeqa
we are able to cover all possibilities for the algebra $\mathfrak{g}$ to be embedded within the $Q$ flux. This matrix $M$ must satisfy the isotropy symmetry and so we have that $\, M = \mathbb{I}_{3} \otimes M_2 \,$, where the four parameters matrix $M_2 \in SL(2,\mathbb{R})$ acts equally in each 2-torus,
\beq
\left(
       \begin{array}{c}
            E^I \\
            \widetilde{E}^I 
       \end{array}
\right)
= \frac{1}{|\Gamma_M |^{2}}
\left(
       \begin{array}{cc}
          -  \alpha  &  \beta \\
          -  \gamma  &\delta   
       \end{array}
\right)
\left(
       \begin{array}{c}
            X^{2I-1} \\
            X^{2I}    
       \end{array}
\right) \ ,
\label{chbasis}
\eeq
for all $I=1,2,3$. Here $|\Gamma_M|=\alpha\delta - \beta\gamma$, and it must be that $|\Gamma_M|\not=0$. In the following, we will refer to the $(\alpha, \beta, \gamma, \delta)$ parameters as the modular parameters.
\\

Applying this, we find the parametrizations presented in \cite{Font:2008vd},

\begin{itemize}
 \item Semisimple $\mathfrak{so(4)}$
\beq
\label{LimC}
\begin{array}{lcl}
c_{0}= \beta\, \delta \, (\beta+\delta) & \quad ; \quad & c_{3}=- \,\alpha\, \gamma \, (\alpha+\gamma) \quad  ,\\
c_{1} = \beta\, \delta \, (\alpha+\gamma) & \quad ; \quad & c_{2}=- \,\alpha\, \gamma \, (\beta+\delta) \quad , \\
\tilde{c}_{2}= \gamma^{2}\, \beta + \alpha^{2}\,\delta & \quad ; \quad & 
\tilde{c}_{1}=- \,(\gamma\, \beta^{2} + \alpha\,\delta^{2}) \quad  .
\end{array}
\eeq

 \item Semisimple $\mathfrak{so(3,1)}$
\beq
\label{LimCSO31}
\begin{array}{lcl}
c_{0}=-\beta  \,\big(\beta ^2+\delta ^2\big) & \quad \ ; \quad &  
c_{3}=\,\alpha \, \big(\alpha ^2+\gamma ^2\big) \quad , \\
c_{1}= -\alpha  \,\big(\beta ^2+\delta ^2\big) & \quad ; \quad & 
c_{2}=\beta \, \big(\alpha ^2+\gamma ^2\big) \quad , \\
\tilde{c}_{2}= -\beta \, (\alpha ^2-\gamma ^2)-2\,\gamma \, \delta  \,\alpha  & \quad ; \quad &
\tilde{c}_{1}=\alpha  \big(\beta ^2-\delta ^2\big) + 2 \,\beta \, \gamma \, \delta \quad  .
\end{array}
\eeq

 \item Non semisimple (ie. direct sum) $\mathfrak{su(2)+u(1)^3}$
\beq
\label{LimCFac}
\begin{array}{lcl}
c_{0}= \beta\, \delta^2  & \quad ; \quad & c_{3}=-\alpha\, \gamma^2\quad , \\
c_{1} = \beta\, \delta \, \gamma & \quad ; \quad & c_{2}= -\alpha\, \gamma \, \delta \quad , \\
\tilde{c}_{2}= \gamma^{2}\, \beta  & \quad ; \quad & 
\tilde{c}_{1}= -\alpha\,\delta^{2} \quad  .
\end{array}
\eeq

 \item Non solvable (ie. semidirect sum) $\mathfrak{iso(3)}$
\beq
\label{LimCFacSemi}
\begin{array}{lcl}
c_{0}=-\delta ^2\,(\beta-\delta) & \quad \ ; \quad &  
c_{3}=\gamma ^2\,(\alpha-\gamma) \quad , \\
c_{1}= -\delta ^2\,(\alpha-\gamma) & \quad ; \quad & 
c_{2}=\gamma ^2\,(\beta-\delta) \quad , \\
\tilde{c}_{2}= \gamma ^2\,(\beta+\delta)- 2\,\gamma \, \delta  \,\alpha  & \quad ; \quad &
\tilde{c}_{1}=-\delta^2\, (\alpha+\gamma)  + 2\, \gamma \, \delta \, \beta \quad  .
\end{array}
\eeq

 \item Solvable (ie. nilpotent) $\mathfrak{nil}$
\beq
\label{LimCNilp}
\begin{array}{lcl}
c_{0}= \delta ^3 & \quad ; \quad & c_{3}=- \gamma^3 \quad  ,\\
c_{1} = \delta^2\, \gamma  & \quad ; \quad & c_{2}=- \delta \, \gamma^2 \quad  ,\\
\tilde{c}_{2}= \delta \, \gamma^{2} & \quad ; \quad & 
\tilde{c}_{1}=- \delta^{2}\, \gamma  \quad .
\end{array}
\eeq
\end{itemize}

It is straightforward to check that these flux configurations satisfy (\ref{BianchiC}). Despite the requirement that the entries in the fluxes are integers, the entries in $M$ are not restricted to being integers. Starting with a configuration where $c_{i} \in \mathbb{Z}$, because the $c_{i}$ have a cubic dependence on the modular parameters, we see that $M'= \sqrt[3]{n} M$ with $n \in \mathbb{Z}$ still gives us $c'_{i}= n \, c_{i} \in \mathbb{Z}$.
\\

When using these parameterisations for the entries in $Q$, the roots of $P_{3}(U)$ can be expressed in terms of the modular parameters and the roots structure, namely the number and type of coincident roots, becomes manifest. Further simplifications can be made by writing the polynomial in terms of the modular variable $\mZ=\frac{\alpha\, U + \beta}{\gamma \, U + \delta}$. This demonstrates that it is not possible to do an $\textrm{SL}(2,\mathbb{Z})$ transformation which alters this,  different algebras lead to different root structures (see table \ref{tableP3roots}).
\\

\begin{table}[htb]
\small{
\renewcommand{\arraystretch}{1.15}
\begin{center}
\begin{tabular}{|c|c|c|}
\hline
Algebra & $\mathcal{P}_3(\mZ) \equiv \frac{P_3(U)}{3\,(\gamma \, U + \delta)^3}$  & Modular roots \\
\hline
\hline
$\mathfrak{so(4)}$ & $\mZ(\mZ+1)$  & $ \mZ= 0 \, \, , \, \,  \infty \,  \, ,\,  \,  -1$        \\
\hline
$\mathfrak{so(3,1)}$ & $-\mZ(\mZ^2+1)$  &   $ \mZ = 0 \, \, , \, \,  +i \,  \, ,\,  \,  -i$ \\
\hline
$\mathfrak{su(2)+u(1)^3}$ & $\mZ$   &   $ \mZ = 0 \, \, , \, \, \infty $ (double) \\
\hline
$\mathfrak{iso(3)}$ & $1-\mZ$     &  $  \mZ=\infty $ (double)  $\,  \, ,\,  \,  +1$\\
\hline
$\mathfrak{nil}$ & $1$  & $ \mZ_{\infty} = \infty$  (triple)\\
\hline
\end{tabular}
\end{center}
\caption{Algebras and their flux-induced polynomials.}
\label{tableP3roots}
}
\end{table}

To analyse this we define the following $2$-dimensional vectors,
\beqa
\label{vectorsZ}
\mZ_0= (\alpha , \beta)            &\hspace{3mm},\hspace{3mm}&  \mZ_{\infty}= (\gamma  , \delta) \ ,\nonumber \\ 
\mZ_{-1}= (\alpha + \gamma , \beta + \delta)        &\hspace{3mm},\hspace{3mm}&   \mZ_{+1}= (\alpha - \gamma , \beta - \delta) \ ,   \\
\mZ_{+i}=  i \left( \sqrt{\alpha ^2+\gamma ^2},\frac{(\alpha  \beta +\gamma  \delta) +i |\Gamma_{M}|}{\sqrt{\alpha^2+\gamma ^2}}   \right)     &\hspace{2mm},\hspace{2mm}&  \mZ_{-i}= i\left(  \sqrt{\alpha ^2+\gamma ^2},\frac{(\alpha  \beta +\gamma  \delta) -i |\Gamma_{M}|}{\sqrt{\alpha^2+\gamma ^2}}   \right),    \nonumber
\eeqa
in such a way that they carry the information about the roots values once they are contracted with $\pmatrix{ U \cr 1 }$. Then the flux-induced polynomial $P_{3}(U)$ for each algebra can be easily reconstructed from its roots structure as
\beqa
\label{prodroots}
P_{3}(U) = 3 \prod_{\Box=roots} \mZ_{\Box}  \pmatrix{ U \cr 1 }\ ,
\eeqa
with $\Box \equiv 0,\infty,-1,+1,+i,-i$ according with the modular roots, as it is shown in table \ref{tableP3roots}. As an example, we reconstruct the cubic $P_{3}(U)$ for the algebra $\mathfrak{so(4)}$. In this case, (\ref{prodroots}) reads
\beqa
P_{3}(U) &=& 3\,  \mZ_{0}  \pmatrix{ U \cr 1 }  \cdot \mZ_{\infty}  \pmatrix{ U \cr 1 } \cdot \mZ_{-1}  \pmatrix{ U \cr 1 } = \nonumber \\
&=&3\,  (\alpha \, U + \beta)\,(\gamma \, U + \delta) \,\, \left[ (\alpha + \gamma) \, U  +  (\beta + \delta) \right] = \nonumber \\
&=&3\,  (\gamma \, U + \delta)^{3} \, \mZ \, (\mZ+1) \ .
\eeqa

Note that $\mathfrak{so(3,1)}$ is unique in the above results, in that it generates a polynomial whose roots are certain to be complex, given the real and non-degenerate nature  of $\Gamma_{M}$.

\subsection{Tadpole cancellation conditions.}

In this IIB orientifold, the Bianchi identities for R-R fluxes can be rephrased as tadpole cancellation conditions for the R-R 4-form $C_4$ and the 8-form $C_8$ which couple to the O3/O7-planes sources allowed by the orientifold group. In the case of $C_{4}$, the flux-induced tadpole arises from the coupling
\beq
\label{c4tad}
\int_{\mathcal{M}_{10}} C_{4} \wedge \bar{H}_{3} \wedge \bar{F}_{3} \ .
\eeq 

The total orientifold charge is -32, due to 64 O3-planes located at the fixed points of the 
$\mathbb{Z}^{3}_{2}$ orientifold involution. Also, it is possible to add D3-branes, of charge $+1$, leading to the cancellation condition
\beq
\label{O3tad}
a_{0}\,b_{3} - a_{1}^{(K)}\,b_{2}^{(K)} + a_{2}^{(K)}\,b_{1}^{(K)} - a_{3}\,b_{0}=N_{3} \ ,
\eeq 
where $N_{3}=32-N_{D3}$ and $N_{D3}$ is the number of D3-branes.
\\

Taking into account the flux-induced tadpole for the $C_{8}$ components of type $C_8 \sim d{\rm vol}_4 \wedge \widetilde{\omega}^I$, where $d{\rm vol}_4$ is the space-time $\mathcal{M}_{4}$ volume 4-form,
\beq
\int_{\mathcal{M}_{10}} C_{8} \wedge (Q \cdot \bar{F}_{3}) \, ,
\label{c8tad}
\eeq 
and expanding the 2-form $(Q \cdot \bar{F}_{3})$ in the basis of 2-forms,
\beq
\label{QFexpan}
( Q \cdot \bar{F}_{3})_{I}=a_{0}\,c_{3}^{(I)}+a_{1}^{(K)}\,C_{2}^{(K I)} - a_{2}^{(K)}\,C_{1}^{(K I)}-a_{3}\,c_{0}^{(I)} \ ,
\eeq
we end up with the three tadpole cancellation conditions
\beq
\label{O7tad}
a_{0}\,c_{3}^{(I)}+a_{1}^{(K)}\,C_{2}^{(K I)} - a_{2}^{(K)}\,C_{1}^{(K I)}-a_{3}\,c_{0}^{(I)}
=N_{7_{I}} \ .
\eeq
Here $\,N_{7_{I}}=-32+N_{D7_I}\,$, where $N_{D7_I}$ is the number of D7-branes which can be added wrapping the $I^{th}$ 4-cycle dual to the 2-torus $\mathbb{T}^{2}_I$.
\\

Going to the isotropic case, this set of conditions reduces to
\beq
\label{O3tadIso}
 a_{0}\,b_{3}-3\,a_{1}\,b_{2}+3\,a_{2}\,b_{1}-a_{3}\,b_{0}=N_{3} \ ,
\eeq
\beq
\label{O7tadIso}
 a_{0}\,c_{3}+a_{1}\,(2\,c_{2} -\tilde{c}_{2})-a_{2}\,(2\,c_{1} -\tilde{c}_{1})-a_{3}\,c_{0}=N_{7} \ .
\eeq

\section{S-duality on top of T-duality.}

S-duality for this orbifold was explored in \cite{Aldazabal:2006up}. All the constraints induced by S-duality, including tadpoles, were derived under the ansatz of systematically applying S-duality transformations to the T-duality invariant constraints. First, we shall focus on the S-dualization of the Bianchi identities (\ref{BianchiT}) and then we will consider the new constraints coming from the S-dualization of the tadpole cancellation conditions.

\subsection{An ansatz for S-duality.}

Applying an S-duality transformation (\ref{QPtransf}) to the non-geometric $Q$ flux, the $QQ=0$ Bianchi identity in (\ref{BianchiT}) gives rise to an $\mathop{\rm SL}(2,\mathbb{Z})_{S}$ triplet of constraints involving the $Q$ and $P$ fluxes, 
\beqa
\label{BianchiPQ} 
Q^{[ab}_{d}Q^{c]d}_{e} = 0 \hspace{5mm},\hspace{5mm} P^{[ab}_{d}P^{c]d}_{e} = 0 \hspace{5mm},\hspace{5mm} Q^{[ab}_{d}P^{c]d}_{e} + P^{[ab}_{d}Q^{c]d}_{e} = 0 \ ,
\eeqa
which, as before, we will denote as $QQ=0$, $PP=0$ and $QP+PQ=0$. In terms of flux entries, the first condition results in that of (\ref{BianchiC}) and the second one reduces to (\ref{BianchiC}) under $Q \to P$, $c \to d$. The third element of the triplet gives the mixing between the $Q$ and $P$ fluxes, which in terms of their entries reduces to
\beqa
\label{QPPQ}
c_{3}d_{0}-c_{2}d_{1}-c_{1}d_{2}+c_{0}d_{3} &=& 0 \ , \nonumber \\ 
c_{1}(d_{1}-\tilde{d}_{1}) + c_{0}(d_{2}-\tilde{d}_{2} )+ d_{0} (c_{2}-\tilde{c}_{2}) +d_{1} (c_{1}-\tilde{c}_{1})&=& 0 \ ,  \\ 
c_{3}(d_{1}-\tilde{d}_{1}) + c_{2} (d_{2}-\tilde{d}_{2}) + d_{2} (c_{2}-\tilde{c}_{2}) + d_{3} (c_{1}-\tilde{c}_{1})&=& 0 \nonumber \ . 
\eeqa

At this point we want to emphasize that the Bianchi constraints (\ref{BianchiPQ}) have been obtained by applying an S-duality transformation to the Bianchi constraint $QQ=0$ of the T-duality invariant effective theory. Starting with an $SL(2,\mathbb{Z})^{7}$-duality invariant algebra \cite{acr} involving the $Q$, $P$, $\bar{H}_{3}$ and $\bar{F}_{3}$ background fluxes, these conditions result slightly modified\footnote{We thank to P. G. C\'amara for discussions on this point.} to $QQ=PP=0$ together with $QP=PQ=0$. However, these modified constraints can be understood as a particular case of (\ref{BianchiPQ}).
\\

We now turn to the second Bianchi constraint on the fluxes, $Q\bar{H}_{3}=0$, and consider what effect S-duality has on it. This constraint equation is extended to mix all of the four fluxes in the IIB with O3/O7-planes space. As derived in \cite{Aldazabal:2006up},
\beq
\label{BianchiQPHF}
Q^{ab}_{[c}\bar{H}_{de]b} = 0 \hspace{3mm}\to\hspace{3mm}  Q^{ab}_{[c}\bar{H}_{de]b} - P^{ab}_{[c}\bar{F}_{de]b} = 0 \ .
\eeq
Once more, we will refer to as $\,Q\bar{H}_{3}-P\bar{F}_{3}=0$. This is an $\mathop{\rm SL}(2,\mathbb{Z})_{S}$ singlet and in terms of the flux entries it reads
\beq
\label{LinBianchi}
\begin{array}{lll}
-c_2 \, b_0 + (c_1-\tilde{c}_1) \, b_1 + c_0 \, b_2 \,\,\,+\,\,\,d_2 \, a_0 - (d_1-\tilde{d}_1) \, a_1 - d_0 \, a_2   &=&0 \ , \nonumber \\
-c_2 \, b_1 + (c_1-\tilde{c}_1) \, b_2 + c_0 \, b_3 \,\,\,+\,\,\,d_2 \, a_1 - (d_1-\tilde{d}_1) \, a_2 - d_0 \, a_3   &=&0 \ , \nonumber \\
-c_3 \, b_0 - (c_2-\tilde{c}_2) \, b_1 + c_1 \, b_2 \,\,\,+\,\,\,d_3 \, a_0 + (d_2-\tilde{d}_2) \, a_1 - d_1 \, a_2   &=&0   \ , \\
-c_3 \, b_1 - (c_2-\tilde{c}_2) \, b_2 + c_1 \, b_3 \,\,\,+\,\,\,d_3 \, a_1 + (d_2-\tilde{d}_2) \, a_2 - d_1 \, a_3   &=&0 \ . \nonumber 
\end{array}
\eeq

\subsection{S-duality and the tadpole cancellation conditions.}

Tadpole constraints were derived in \cite{Aldazabal:2006up} for the full non-isotropic case. There are two kinds of flux-induced tadpoles: tadpoles which exist in an $\mathop{\rm SL}(2,\mathbb{Z})_ {S}$ triplet \cite{Bergshoeff:2006ic} and arise from the $(Q \cdot \bar{F}_{3})$ tadpole in (\ref{c8tad}), 
\beqa
\int_{\mathcal{M}_{10}}  C_{8} \wedge (Q \cdot \bar{F}_{3}) \hspace{5mm},\hspace{5mm} \int_{\mathcal{M}_{10}}  \tilde{C}_{8} \wedge (P \cdot \bar{H}_{3}) \hspace{5mm},\hspace{5mm} \int_{\mathcal{M}_{10}}  C'_{8} \wedge (Q \cdot \bar{H}_{3} + P \cdot \bar{F}_{3}) \ ,
\eeqa
and the singlet tadpole in (\ref{c4tad}) which remains unchanged from the T-duality invariant effective theory.
\\

Using the expansions
\beqa
(P \cdot \bar{H}_{3})_{I}&=&  b_{0}\,d_{3}^{(I)}+b_{1}^{(K)}\,\mathcal{D}_{2}^{(K I)} - b_{2}^{(K)}\,\mathcal{D}_{1}^{(K I)}-b_{3}\,d_{0}^{(I)} \ ,\\
(Q \cdot \bar{H}_{3})_{I}&=&  b_{0}\,c_{3}^{(I)}+b_{1}^{(K)}\,\mathcal{C}_{2}^{(K I)} - b_{2}^{(K)}\,\mathcal{C}_{1}^{(K I)}-b_{3}\,c_{0}^{(I)}  \ ,\\
(P \cdot \bar{F}_{3})_{I}&=&  a_{0}\,d_{3}^{(I)}+a_{1}^{(K)}\,\mathcal{D}_{2}^{(K I)} - a_{2}^{(K)}\,\mathcal{D}_{1}^{(K I)}-a_{3}\,d_{0}^{(I)} \ ,
\eeqa
the new tadpole cancellation conditions for the $\tilde{C}_8$ and $C'_8$ potentials read
\beq
\label{NS7tad}
b_{0}\,d_{3}^{(I)}+b_{1}^{(K)}\,\mathcal{D}_{2}^{(K I)} - b_{2}^{(K)}\,\mathcal{D}_{1}^{(K I)}-b_{3}\,d_{0}^{(I)}
=\tilde{N}_{7_{I}}
\eeq
and
\beqa
\label{I7tad}
b_{0}\,c_{3}^{(I)}&+&b_{1}^{(K)}\,\mathcal{C}_{2}^{(K I)} - b_{2}^{(K)}\,\mathcal{C}_{1}^{(K I)}-b_{3}\,c_{0}^{(I)} + \nonumber \\
a_{0}\,d_{3}^{(I)}&+&a_{1}^{(K)}\,\mathcal{D}_{2}^{(K I)} - a_{2}^{(K)}\,\mathcal{D}_{1}^{(K I)}-a_{3}\,d_{0}^{(I)}=N'_{7_{I}}
\eeqa
respectively, where $\tilde{N}_{7_{I}}$ accounts for the number of NS7-branes and $N'_{7_{I}}$ is related to the number of I7-branes (bound states \cite{Bergshoeff:2006ic} of D7 and NS7 branes) which can be added to the system wrapping the $I^{th}$ 4-cycle dual to the 2-torus $\mathbb{T}^{2}_I$.
\\

Restricting ourselves to the case of isotropic fluxes, these conditions simplify to
\beq
\label{NS7tadIso}
 b_{0}\,d_{3}+b_{1}\,(2\,d_{2} -\tilde{d}_{2})-b_{2}\,(2\,d_{1} -\tilde{d}_{1})-b_{3}\,d_{0}=\tilde{N}_{7}
\eeq
and
\beqa
\label{I7tadIso}
&& b_{0}\,c_{3}+b_{1}\,(2\,c_{2} -\tilde{c}_{2})-b_{2}\,(2\,c_{1} -\tilde{c}_{1})-b_{3}\,c_{0} + \nonumber \\
&& a_{0}\,d_{3}+a_{1}\,(2\,d_{2} -\tilde{d}_{2})-a_{2}\,(2\,d_{1} -\tilde{d}_{1})-a_{3}\,d_{0} = N'_{7} \ .
\eeqa
A further simplification can be made, as noted in \cite{Aldazabal:2006up}, 
\begin{eqnarray}
\label{implication}
Q\bar{H}_{3}=0 \Rightarrow Q \cdot \bar{H}_{3}=0 \hspace{5mm},\hspace{5mm} P\bar{F}_{3}=0 \Rightarrow P \cdot \bar{F}_{3}=0 \ .
\end{eqnarray}

\section{The non-geometric background fluxes.}

In this section, we try to clarify the role played by the non-geometric $P$ flux in terms of deformations of Lie algebras. Before considering the algebraic problem of solving the constraints $PP=0$ and $QP+PQ=0$, we focus our attention on understanding the problem from a different point of view, that of the effect of the $P$ flux over the T-duality invariant gauge subalgebra generated by the $Q$ flux.

\subsection{A note on deformations of Lie algebras.}

To start with, we present a brief introduction to the topic of deformations of Lie algebras, in which we take the notation and conventions from \cite{CampoamorStursberg}. Let us start with a general Lie algebra $\mathcal{L}$ defined by its brackets\footnote{We define generators with an upper index in analogy with the commutation relations $[X^{a},X^{b}]=Q^{ab}_{c}\, X^{c}$ we are dealing with.}
\beq
[X^{a},X^{b}]=C^{ab}_{c}\, X^{c} \ .
\eeq

These relations define an algebra iff Jacobi identities are fulfilled, namely $C^{[ab}_{e} \, C^{c]e}_{d}=0$. For our purposes, it will be interesting to define the second cohomology class of the algebra, $H^{2}(\mathcal{L},\mathcal{L})$. It contains 2-cocycles $\varphi \in H^{2}(\mathcal{L},\mathcal{L})$ that are closed under the action of an exterior derivation $d$ without being coboundaries. More formally, a cocycle $\varphi \in H^{2}(\mathcal{L},\mathcal{L})$ is a bilinear antisymmetric form that satisfies the constraint
\beqa 
\label{cohomology}
d \varphi (X^{a},X^{b},X^{c}) &:=& [X^{a},\varphi(X^{b},X^{c})] + [X^{c},\varphi(X^{a},X^{b})] + [X^{b},\varphi(X^{c},X^{a})] +  \\
&+& \varphi(X^{a},[X^{b},X^{c}]) + \varphi(X^{c},[X^{a},X^{b}]) + \varphi(X^{b},[X^{c},X^{a}]) = 0 \ , \nonumber
\eeqa
for any $X^{a}$, $X^{b}$, $X^{c}$ of $\mathcal{L}$.
\\

Moreover, for $\varphi$ to define a deformation of $\mathcal{L}$ that is also a Lie algebra, i.e. it also satisfies the new Jacobi identities, an additional integrability condition has to be imposed. The 2-cocycle $\varphi$ is integrable if it satisfies
\beq
\label{integrability}
\varphi(\varphi(X^{a},X^{b}),X^{c}) + \varphi(\varphi(X^{c},X^{a}),X^{b}) + \varphi(\varphi(X^{b},X^{c}),X^{a}) = 0 \ .
\eeq

If both conditions, named the cohomology and the integrability conditions, are fulfilled then the linear deformation $\mathcal{L}+\varphi$ is also a Lie algebra \cite{Weimar-Woods}, which we will denote as $\mathcal{L}_{\varphi}$ with the deformed bracket
\beq
[X^{a},X^{b}]_{\varphi}=C^{ab}_{c}\, X^{c} + \varphi(X^{a},X^{b}) \ .
\eeq

In particular, nullity of $H^{2}(\mathcal{L},\mathcal{L})$ implies that any deformation $\mathcal{L}_{\varphi}$ is isomorphic to $\mathcal{L}$ and in that case $\mathcal{L}$ is called rigid or stable. However, in general, $\mathcal{L}_{\varphi}$ and $\mathcal{L}$ are not isomorphic.
\\

To clarify the utility of deformed Lie algebras in the problem of S-duality and non-geometric fluxes, let us consider a deformation $\varphi(X^{a},X^{b}):= \alpha^{ab}_{c}\,X^{c}$ with $\alpha^{ab}_{c}=-\alpha^{ba}_{c}$, so the cohomology condition (\ref{cohomology}) can be rewritten as
\beq
\label{cohomology2}
C^{[ab}_{e}\,\alpha^{c]e}_{d} + \alpha^{[ab}_{e}\,C^{c]e}_{d} = 0 \ ,
\eeq
while the integrability condition results in
\beq
\label{integrability2}
\alpha^{[ab}_{e}\,\alpha^{c]e}_{d} = 0 \ .
\eeq

At this point, the role of the non-geometric $P$ flux becomes clear by identifying $C^{ab}_{c}=Q^{ab}_{c}$ and $\alpha^{ab}_{c}=P^{ab}_{c}$. The non-geometric $Q$ flux defines the gauge subalgebra of the T-duality invariant effective theory while the non-geometric $P$ flux can be implemented as deformations of this subalgebra by an element of its second cohomology class. The $PP=0$ and $QP+PQ=0$ constraints in (\ref{BianchiPQ}) are simply the integrability (\ref{integrability2}) and cohomology (\ref{cohomology2}) conditions for the non-geometric $P$ flux to define such deformations. The T-duality invariant gauge subalgebra is trivially recovered when the deformation vanishes, i.e. $P=0$, and just the original condition $QQ=0$ remains unchanged. Another possibility to recover it is to fix $P=Q$, which is related to the $P=0$ case by the $SL(2,\mathbb{Z})_{S}$ transformation $\Theta_{S}=\pmatrix{0 & -1 \cr 1 & -1}$. This can be interpreted as a deformation of the T-duality invariant gauge subalgebra by itself.

\subsection{Solving the integrability condition.}

The integrability condition $PP=0$ is straightforwardly solved by imposing that $P$ becomes the structure constants of a Lie algebra $\mathfrak{g}_{P}$ belonging to the set of non trivial six dimensional Lie algebras compatibles with the orbifold symmetries, in analogy with the $QQ=0$ condition. To solve both the $QQ=0$ and $PP=0$ conditions simultaneously, we pick two algebras, $\mathfrak{g}_{Q}$ and $\mathfrak{g}_{P}$, from table \ref{canonicQalgebras} and equate the $Q$ and $P$ fluxes to the transformed $g_Q$ and $g_P$ structure constants, 
\beq
Q=M^{-1}_{Q} \, M^{-1}_{Q} \, g_{Q} \,\, M_{Q} \hspace{3mm},\hspace{3mm} P=M^{-1}_{P} \, M^{-1}_{P} \, g_{P} \,\, M_{P} \ ,
\label{LimQP}
\eeq
with specific modular parameters
\beqa
\Gamma_{Q} = \pmatrix{\alpha_{q} & \beta_{q} \cr \gamma_{q} & \delta_{q}} \hspace{3mm}\textrm{and}\hspace{4mm} \Gamma_{P} = \pmatrix{\alpha_{p} & \beta_{p} \cr \gamma_{p} & \delta_{p}} \ ,
\label{ModularQP}
\eeqa
entering in $M_{Q}$ and $M_{P}$ defined as in (\ref{chbasis}). We end up with a general parametrization of the non-geometric fluxes, ie.  $Q = Q(\alpha_{q},\beta_{q},\gamma_{q},\delta_{q})$ and $P = P(\alpha_{p},\beta_{p},\gamma_{p},\delta_{p})$, analogous to those of (\ref{LimC})-(\ref{LimCNilp}).
\\

Recalling (\ref{ModularQP}), we can now define the two modular variables
\beq
\label{modularZQZP}
\mZ_Q=\frac{\alpha_{q} \, U + \beta_{q}}{\gamma_{q} \, U + \delta_{q}} \hspace{6mm},\hspace{6mm} \mZ_P=\frac{\alpha_{p} \, U + \beta_{p}}{\gamma_{p} \, U + \delta_{p}} \ .
\eeq
Expressing the superpotential polynomials due to $Q$ and $P$ in terms of these, we have $\mathcal{P}_{3}(\mZ_{Q}) \equiv P_{3}(U)/3(\gamma_{q} \, U + \delta_{q})^3$ and $\mathcal{P}_{4}(\mZ_{P}) \equiv -P_{4}(U)/3(\gamma_{p} \, U + \delta_{p})^3$, where the polynomials relating to $\mathfrak{g}_{Q}$ and $\mathfrak{g}_{P}$ can be simply read off from table \ref{tableP3roots}, upon replacing $\mZ$ by $\mZ_Q$ and $\mZ_P$ respectively.

\subsection{Solving the cohomology condition.}

In terms of the flux entries, the cohomology $QP+PQ=0$ constraints are those of (\ref{QPPQ}). Since the expressions for the entries of $Q$ and $P$ are in terms of the modular parameters the cohomology condition puts constraints on their possible values. Finding the space of valid flux entries is difficult because the constraints are polynomials in terms of the $8$ modular parameters. However, these polynomials form the generators of the ideal $\langle QP+PQ \rangle$ in the ring of polynomials $\mathbb{C}[\alpha_{q},\ldots,\delta_{p}]$ and so we can use an algebraic geometry method of prime decomposition to split $\langle QP+PQ \rangle$ into its prime ideals, $J_{i}$. One such method is the GTZ algorithm, which is implemented within \textit{Singular}. Each prime ideal has a solution space, the variety $\mathbb{V}_{i}$, which is a subset of $\mathbb{V}$, the variety of $\langle QP+PQ \rangle$ and because we are working with prime ideals, their varieties do not intersect other than at a finite number of disjoint points. Therefore, given the decomposition
\begin{eqnarray}
\langle  \; QP+PQ \; \rangle = J_{1} \cap \ldots \cap J_{n} \ ,
\end{eqnarray}
in order to satisfy $QP+PQ=0$, we need only to solve the set of equations $f_{i,j}=0$, where $J_{i}= \langle  \; f_{i,1},f_{i,2},\ldots,f_{i,m} \; \rangle$, though to completely account for all possible solutions each prime ideal must be analysed. An ideal automatically has at least one prime ideal but in the case of some of the $(\mathfrak{g}_{Q},\mathfrak{g}_{P})$ pairings, we find as many as three prime ideals of varying complexity. These relate the $\Gamma_{Q}$ and $\Gamma_{P}$ modular matrices and so restrict the transformations which are needed to bring the $Q$ and $P$ fluxes (understood as structure constants) to their canonical form.
\\

For purpose of illustration we consider the example $\mathfrak{g}_{Q} = \mathfrak{su(2)} + \mathfrak{u(1)^{3}}$ and $\mathfrak{g}_{P} = \mathfrak{so(4)}$. We have individual parametrization of the following format, 

\begin{itemize}

\item $Q$ flux fixing the gauge subalgebra in the T-duality invariant supergravity to be $\mathfrak{g}_{Q} = \mathfrak{su(2)} + \mathfrak{u(1)^{3}}$,
\beq
\label{LimCexample}
\begin{array}{lcl}
c_{0}= \beta_{q}\, \delta_{q}^2  & \quad ; \quad & c_{3}=-\alpha_{q}\, \gamma_{q}^2 \ , \\
c_{1} = \beta_{q}\, \delta_{q} \, \gamma_{q} & \quad ; \quad & c_{2}= -\alpha_{q}\, \gamma_{q} \, \delta_{q} \ , \\
\tilde{c}_{2}= \gamma_{q}^{2}\, \beta_{q}  & \quad ; \quad & 
\tilde{c}_{1}= -\alpha_{q}\,\delta_{q}^{2} \ . 
\end{array}
\eeq

\item $P$ flux fixing the original gauge subalgebra in the T-duality invariant supergravity to be deformed by $\mathfrak{g}_{P} = \mathfrak{so(4)}$,
\beq
\label{LimDexample}
\begin{array}{lcl}
d_{0}= \beta_{p}\, \delta_{p} \, (\beta_{p}+\delta_{p}) & \quad ; \quad & d_{3}=- \,\alpha_{p}\, \gamma_{p} \, (\alpha_{p}+\gamma_{p}) \quad , \\
d_{1} = \beta_{p}\, \delta_{p} \, (\alpha_{p}+\gamma_{p}) & \quad ; \quad & d_{2}=- \,\alpha_{p}\, \gamma_{p} \, (\beta_{p}+\delta_{p}) \quad , \\
\tilde{d}_{2}= \gamma_{p}^{2}\, \beta_{p} + \alpha_{p}^{2}\,\delta_{p} & \quad ; \quad & 
\tilde{d}_{1}=- \,(\gamma_{p}\, \beta_{p}^{2} + \alpha_{p}\,\delta_{p}^{2}) \ . 
\end{array}
\eeq

\end{itemize}

This leads to a $\langle QP+PQ  \rangle$ cohomology condition ideal which has three prime ideals in its decomposition,
\beqa
\label{idealprimfacexample}
J_{1} & = & \langle \; \alpha _q \beta _p-\beta _q \alpha _p   \,\,,\,\, \gamma _q \delta _p-\delta _q \gamma _p   \; \rangle  \ ,\nonumber \\
J_{2} & = & \langle \; \alpha _q \delta _p-\beta _q \gamma _p  \,\,\,\,,\,\, \gamma _q \beta _p -\delta _q \alpha _p      \; \rangle  \ ,\\
J_{3} & = & \langle \;  \gamma _q  (\beta _p +\delta _p) - \delta _q (\alpha _p + \gamma _p)        \; \rangle \ .  \nonumber
\eeqa

These constraints can be rewritten in terms of entries in $2$ dimensional vectors
\begin{eqnarray}
\mathbf{u} = \pmatrix{u_{1}\cr u_{2}} \quad , \quad \mathbf{v} = \pmatrix{v_{1}\cr v_{2}} \quad \Rightarrow \quad u_{1} v_{2} - u_{2} v_{1} = 0 \quad \Leftrightarrow \quad \mathbf{u}\times \mathbf{v} = 0 \ .
\end{eqnarray} 
If two vectors satisfy $\mathbf{u}\times \mathbf{v} = 0$ then they are parallel, which we denote by $\mathbf{u}\parallel\mathbf{v}$. With this notation and using the vectors given in (\ref{vectorsZ}), the cohomology condition becomes
\beq
\label{2cocycleexample}
\begin{array}{ccccc}
J_{1} & = & \langle \;    \mZ^{Q}_{0} \times \mZ^{P}_{0} \,\,\,,\,\,\, \mZ^{Q}_{\infty} \times \mZ^{P}_{\infty}   \; \rangle  & \Leftrightarrow & \,\, \mZ^{Q}_{0} \parallel \mZ^{P}_{0} \,\,\,,\,\,\, \mZ^{Q}_{\infty} \parallel \mZ^{P}_{\infty} \ .\nonumber \\
J_{2} & = & \langle \;   \mZ^{Q}_{0} \times \mZ^{P}_{\infty} \,\,\,,\,\,\, \mZ^{Q}_{\infty} \times \mZ^{P}_{0}     \; \rangle & \Leftrightarrow & \,\, \mZ^{Q}_{0} \parallel \mZ^{P}_{\infty} \,\,\,,\,\,\, \mZ^{Q}_{\infty} \parallel \mZ^{P}_{0} \ .\\
J_{3} & = & \langle \;   \mZ^{Q}_{\infty} \times \mZ^{P}_{-1}     \; \rangle & \Leftrightarrow & \,\, \mZ^{Q}_{\infty} \parallel \mZ^{P}_{-1} \ . \nonumber
\end{array}
\eeq
In each case the prime ideal's generating functions can be rewritten as a vanishing cross product. Infact, this happens for all prime ideals of all possible pairings $(\mathfrak{g}_Q,\mathfrak{g}_P)$.  Therefore, the prime ideals of $\langle QP+PQ \rangle$ can be viewed as geometric constraints on the position of the vectors representing the roots of the cubic polynomials $P_{3}(U)$ and $P_{4}(U)$. Specifically, when the polynomials themselves are computed, this is equivalent to $P_{3}(U)$ and $P_{4}(U)$ sharing some roots.
\\

It is worth noticing here that the $J_1=0$ and $J_2=0$ solutions also imply the piecewise vanishing $QP=PQ=0$, unlike $J_{3}=0$. Moreover, $J_1=0$ can be translated into $\mZ_{P} \propto \mZ_{Q}$ while $J_2=0$ implies $\mZ_{P} \propto \mathcal{S} \mZ_{Q}$, where $\mathcal{S}$ is the inversion generator of $SL(2,\mathbb{Z})$.
\\

The full list of the vector alignments arising from the different prime ideals of the cohomology condition are given in table \ref{tablegQgP} for each algebra pairing $(\mathfrak{g}_Q,\mathfrak{g}_P)$. Most of these solutions (those labelled by $(*)$) disappear under the more restrictive condition $QP=PQ=0$, or equivalently, not all the pairings are allowed in the $SL(2,\mathbb{Z})^{7}$-duality invariant theory. Apart from each algebra being deformed by itself, there are the following possibilities in an $SL(2,\mathbb{Z})^{7}$-duality invariant supergravity: $\mathfrak{so(4)}$ can be deformed by $\mathfrak{su(2)+u(1)^{3}}$ ; $\mathfrak{su(2)+u(1)^{3}}$ can be deformed by $\mathfrak{so(4)}$ and by $\mathfrak{nil}$ ;  $\mathfrak{iso(3)}$ can be deformed by $\mathfrak{nil}$ and $\mathfrak{nil}$ can be deformed by $\mathfrak{su(2)+u(1)^{3}}$ and by $\mathfrak{iso(3)}$.
\\

\begin{table}[htb]
\small{
\renewcommand{\arraystretch}{1.15}
\begin{center}
\begin{scriptsize}
\begin{tabular}{c|c|c|c|c|c}
  & \multicolumn{5}{c}{$\mathfrak{g}_{P}$ deformation} \\
\\

$\mathfrak{g}_{Q}$ original & $\mathfrak{so(4)}$  & $\mathfrak{so(3,1)}$  &     $\mathfrak{su(2)+u(1)^3}$ &    $\mathfrak{iso(3)}$ & $\mathfrak{nil}$  \\

& & & & & \\

\hline

   &    $\mZ^{Q}_{0} \parallel \mZ^{P}_{0} \,,\, \mZ^{Q}_{\infty} \parallel \mZ^{P}_{\infty}$     &  &        $\mZ^{Q}_{0} \parallel \mZ^{P}_{0} \,,\, \mZ^{Q}_{\infty} \parallel \mZ^{P}_{\infty}$&  & \\
$\mathfrak{so(4)}$  &   $\mZ^{Q}_{0} \parallel \mZ^{P}_{\infty} \,,\, \mZ^{Q}_{\infty} \parallel \mZ^{P}_{0}$ &   $\mZ^{Q}_{-1} \parallel \mZ^{P}_{0} \, (*)$  &   $\mZ^{Q}_{0} \parallel \mZ^{P}_{\infty} \,,\, \mZ^{Q}_{\infty} \parallel \mZ^{P}_{0}$&  $\mZ^{Q}_{-1} \parallel \mZ^{P}_{+1}\, (*)$ &   $\mZ^{Q}_{-1} \parallel \mZ^{P}_{\infty}\, (*)$ \\
     &    $\mZ^{Q}_{-1} \parallel \mZ^{P}_{-1}\, (*)$   &  &        $\mZ^{Q}_{-1} \parallel \mZ^{P}_{\infty}\, (*)$  & &  \\
\hline
 & & $\mZ^{Q}_{+i} \parallel \mZ^{P}_{+i} \,,\, \mZ^{Q}_{-i} \parallel \mZ^{P}_{-i}$ &  &  & \\  
$\mathfrak{so(3,1)}$ &  $\mZ^{Q}_{0} \parallel \mZ^{P}_{-1}\, (*)$ &   $\mZ^{Q}_{+i} \parallel \mZ^{P}_{-i} \,,\, \mZ^{Q}_{-i} \parallel \mZ^{P}_{+i}$   & $\mZ^{Q}_{0} \parallel \mZ^{P}_{\infty}\, (*)$ & $\mZ^{Q}_{0} \parallel \mZ^{P}_{+1}\, (*)$& $\mZ^{Q}_{0} \parallel \mZ^{P}_{\infty}\, (*)$   \\
 & & $\mZ^{Q}_{0} \parallel \mZ^{P}_{0}\, (*)$ &  &  & \\  

\hline

 &$\mZ^{Q}_{0} \parallel \mZ^{P}_{0} \,,\, \mZ^{Q}_{\infty} \parallel \mZ^{P}_{\infty}$ & &  $\mZ^{Q}_{0} \parallel \mZ^{P}_{\infty} \,,\, \mZ^{Q}_{\infty} \parallel \mZ^{P}_{0}$ & & \\  
$\mathfrak{su(2)+u(1)^3}$& $\mZ^{Q}_{0} \parallel \mZ^{P}_{\infty} \,,\, \mZ^{Q}_{\infty} \parallel \mZ^{P}_{0}$ & $\mZ^{Q}_{\infty} \parallel \mZ^{P}_{0}\, (*)$ &  & $\mZ^{Q}_{\infty} \parallel \mZ^{P}_{+1}\, (*)$ &  $\mZ^{Q}_{\infty} \parallel \mZ^{P}_{\infty}$ \\
 & $\mZ^{Q}_{\infty} \parallel \mZ^{P}_{-1}\, (*) $ & & $\mZ^{Q}_{\infty} \parallel \mZ^{P}_{\infty} $ &  & \\

\hline

 & &  &  & $\mZ^{Q}_{\infty} \parallel \mZ^{P}_{\infty} $ & $\mZ^{Q}_{\infty} \parallel \mZ^{P}_{\infty} $ \\  
$\mathfrak{iso(3)}$ &  $\mZ^{Q}_{+1} \parallel \mZ^{P}_{-1} \, (*)$ & $\mZ^{Q}_{+1} \parallel \mZ^{P}_{0}\, (*)$ &   $\mZ^{Q}_{+1} \parallel \mZ^{P}_{\infty}\, (*) $ &  & \\
 & &  &  & $\mZ^{Q}_{+1} \parallel \mZ^{P}_{+1} \, (*)$ & $\mZ^{Q}_{+1} \parallel \mZ^{P}_{\infty} \, (*)$ \\

\hline

 & &  &  & $\mZ^{Q}_{\infty} \parallel \mZ^{P}_{\infty} $ & \\  
$\mathfrak{nil}$ &  $\mZ^{Q}_{\infty} \parallel \mZ^{P}_{-1} \, (*)$ & $\mZ^{Q}_{\infty} \parallel \mZ^{P}_{0}\, (*)$ &   $\mZ^{Q}_{\infty} \parallel \mZ^{P}_{\infty} $ &  &  $\mZ^{Q}_{\infty} \parallel \mZ^{P}_{\infty} $\\
 & &  &  & $\mZ^{Q}_{\infty} \parallel \mZ^{P}_{+1} \, (*)$ & \\  

\end{tabular}
\end{scriptsize}
\end{center}
\caption{Cohomology condition in terms of the root alignments. The branches labelled by $(*)$ disappear under the more restrictive condition $QP=PQ=0$. Under the inversion $\,S \rightarrow -1/S\,$ transformation, the algebras $\mathfrak{g}_{Q}$ and $\mathfrak{g}_{P}$ are exchanged resulting in the symmetry of this table (and all the forthcoming ones).}
\label{tablegQgP}
}
\end{table}

We note how in several $(\mathfrak{g}_{Q},\mathfrak{g}_{P})$ pairings there are two, even three, different ways (branches) to solve the cohomology condition. In section 7 we provide an example based on $\mathfrak{g}_{Q} = \mathfrak{su(2)} + \mathfrak{u(1)^{3}}$ and $\mathfrak{g}_{P} =\mathfrak{so(4)}$ for which supersymmetric Minkowski vacua only exist in one of these branches, i.e. $\mZ^{Q}_{\infty} \parallel \mZ^{P}_{-1} $. However, supersymmetric AdS$_{4}$ solutions can be found in the other branches, i.e. $\mZ^{Q}_{0} \parallel \mZ^{P}_{0}$ together with $\mZ^{Q}_{\infty} \parallel \mZ^{P}_{\infty}$.

\section{The $\bar{H}_3$ and $\bar{F}_3$ background fluxes.}

Now, we shall proceed to solve the set of constraints (\ref{LinBianchi}) coming from the SL$(2,\mathbb{Z})_{S}$ singlet Bianchi equation. Schematically,  these constraints can be written as a linear system
\beqa
\label{Phisystem}
({\Phi_{Q}})_{i}^{\,\,j}  \,  b_{j} = ({\Phi_{P}})_{i}^{\,\,j} \, a_{j} \ ,
\eeqa
where ${\Phi_{Q}}$ and ${\Phi_{P}}$ are $4\times4$ rank two matrices depending on the non-geometric $Q$ and $P$ fluxes respectively.
\\

Since modular variables are more transparent to work with, we decide to use a universal parametrization for NS-NS and R-R fluxes based on the complete decomposition
\beq
\label{P2P1expansion}
P_{2}(U)=(\gamma_{q} \, U + \delta_{q})^{3} \,  \mathcal{P}_{2}(\mZ_{Q}) \hspace{5mm},\hspace{5mm} P_{1}(U)=-(\gamma_{p} \, U + \delta_{p})^{3} \, \mathcal{P}_{1}(\mZ_{P}) \ ,
\eeq
with $\,\,\mathcal{P}_{2}(\mZ_{Q})=\sum_{i=0}^{3} \epsilon_{i}\, \mZ_{Q}^{i}\,\,$ and $\,\,\mathcal{P}_{1}(\mZ_{P})=\sum_{i=0}^{3} \rho_{i}\, \mZ_{P}^{i}\,\,$. Under this decomposition, the NS-NS $\bar{H}_{3}$ flux entries are parametrised as
\beq
\label{LimB}
\\
\left(
\begin{array}{c}
b_0 \\
b_1 \\
b_2 \\
b_3
\end{array}
\right)
=
\left(
\begin{array}{cccc}
 -\beta _q^3 & -\beta _q \delta _q^2 & -\beta _q^2 \delta _q & -\delta _q^3 \\
 \alpha _q \beta _q^2 & \frac{1}{3} \delta _q \left(2 \beta _q \gamma _q+\alpha _q \delta _q\right) & \frac{1}{3} \beta _q \left(\beta _q
   \gamma _q+2 \alpha _q \delta _q\right) & \gamma _q \delta _q^2 \\
 -\alpha _q^2 \beta _q & -\frac{1}{3} \gamma _q \left(\beta _q \gamma _q+2 \alpha _q \delta _q\right) & -\frac{1}{3} \alpha _q \left(2 \beta
   _q \gamma _q+\alpha _q \delta _q\right) & -\gamma _q^2 \delta _q \\
 \alpha _q^3 & \alpha _q \gamma _q^2 & \alpha _q^2 \gamma _q & \gamma _q^3
\end{array}
\right)
\left(
\begin{array}{c}
\epsilon_0 \\
\epsilon_1 \\
\epsilon_2 \\
\epsilon_3
\end{array}
\right)
\\
\eeq
and those for R-R $\bar{F}_{3}$ flux, $a_i$, have the same form\footnote{These universal parametrizations are well defined because their Jacobians  have determinants $-|\Gamma_{Q}|^{6}/9$ and $-|\Gamma_{P}|^{6}/9$ so they never vanish, provided the isomorphisms used for bringing non-geometric fluxes to their canonical form are not singular.} upon replacing the subscript $q \to p$ and $\epsilon_i \to \rho_i$.
\\

Fixing a pairing $(\mathfrak{g}_{Q},\mathfrak{g}_{P})$ and substituting (\ref{LimQP}) and (\ref{LimB}) into (\ref{Phisystem}), we obtain
\beqa
({\tilde{\Phi}_{Q}})_{i}^{\,\,j}  \,  \epsilon_{j} = ({\tilde{\Phi}_{P}})_{i}^{\,\,j} \, \rho_{j} \ ,
\label{Phisystem_epsi}
\eeqa
where $\tilde{\Phi}_{Q}$ and $\tilde{\Phi}_{P}$ depend on the modular matrices $\Gamma_{Q}$ and $\Gamma_{P}$ defined in (\ref{ModularQP}). Both ${\tilde{\Phi}_{Q}}$ and ${\tilde{\Phi}_{P}}$ are linear transformations and therefore the solutions space of (\ref{Phisystem_epsi}) can be obtained from the intersection of their images 
\beq
\mathfrak{I}_{QP} \equiv \mathfrak{Im}(\tilde{\Phi}_{Q}) \cap \mathfrak{Im}(\tilde{\Phi}_{P})\ .
\label{IPQ}
\eeq
The parameters $\epsilon_{i}$ and $\rho_{j}$ belong to the $\tilde{\Phi}_{Q}$ and $\tilde{\Phi}_{P}$ antimages of $\mathfrak{I}_{QP}$ respectively,
\beq
\begin{array}{lll}
\vec{\epsilon} &\in & \tilde{\Phi}_{Q}^{-1}(\mathfrak{I}_{QP}) \ ,\\
\vec{\rho} &\in & \tilde{\Phi}_{P}^{-1}(\mathfrak{I}_{QP}) \ .
\label{Limepsirho}
\end{array}
\eeq

Therefore we denote a background for the $\bar{H}_{3}$ and $\bar{F}_{3}$ fluxes solving (\ref{Phisystem_epsi}), by a pair of vectors $(\, \vec{\epsilon},\vec{\rho} \,)$ satisfying (\ref{Limepsirho}). The main features of this background, such as its dimension or its flux-induced $C'_{8}$ tadpole, are severely restricted by the non-geometric background we have previously imposed. Furthermore, we are able to distinguish between two non-geometric flux configurations by seeing whether or not $\mathcal{I}_{QP}$ becomes trivial.
\begin{itemize}

\item \underline{Non-geometric type A configuration:} A non-geometric background satisfying 
\beq
\label{IQPeq0}
\mathfrak{I}_{QP} =  \left\lbrace \textbf{0} \right\rbrace \ ,
\eeq
fixes the NS-NS and R-R background fluxes to be  $\vec{\epsilon} \in \mathrm{ker}(\tilde{\Phi}_{Q})$ $(Q\bar{H}_{3}=0)$ and $\vec{\rho} \in \mathrm{ker}(\tilde{\Phi}_{P})$ $(P\bar{F}_{3}=0)$. This has dimension 4 and, according to (\ref{implication}), does not generate a flux-induced $C'_{8}$ tadpole,
\beqa
N'_{7}=0 \hspace{1cm} \mathrm{(type \, A)}.
\eeqa

\item \underline{Non-geometric type B configuration:} A non-geometric background satisfying
\beq
\label{IQPneq0}
\mathfrak{I}_{QP} \neq  \left\lbrace \textbf{0} \right\rbrace \ ,
\eeq
results in a less restricted one for the NS-NS and R-R fluxes. It is a 6 dimensional background for which a flux-induced $C'_{8}$ tadpole can be generated. This can always be written as
\beqa
\label{TadI7Delta}
N'_{7} = \Delta_{Q} \,\, |\Gamma_Q|^3 + \Delta_{P} \,\, |\Gamma_P|^3 \hspace{1cm} \mathrm{(type \, B)} \ ,
\eeqa
with $\Delta_{Q}$ and $\Delta_{P}$ depending on $\epsilon_i$ and $\rho_i$ respectively\footnote{$\mathrm{ker}(\tilde{\Phi}_{Q})$, $\mathrm{ker}(\tilde{\Phi}_{P})$, $\Delta_{Q}$ and $\Delta_{P}$ differ for each pairing $(\mathfrak{g}_{Q},\mathfrak{g}_{P})$, being easily computed in each case.} and vanishing in the special case of $\vec{\epsilon} \in \mathrm{ker}(\tilde{\Phi}_{Q})$ and $\vec{\rho} \in \mathrm{ker}(\tilde{\Phi}_{P})$.
\end{itemize}

Let us explain a little bit more about the preceding classification. Starting with a non-geometric background for the $Q$ and $P$ fluxes, that satisfies both the integrability and the cohomology conditions, it will be either a type A or a type B configuration. For this to be a type B it has to fulfil (\ref{IQPneq0}), which can be rephrased as a single roots alignment, as shown in table \ref{tablePhisystem} for all the possible pairings $(\mathfrak{g}_{Q},\mathfrak{g}_{P})$. If the non-geometric background we are working with generates that alignment, we are dealing with a type B configuration. Otherwise it is type A configuration. This is determined by the way (branch) we followed for solving the cohomology condition (see table \ref{tablegQgP}). 
\\

To illustrate this, we consider an example where $\mathfrak{g}_{Q} = \mathfrak{su(2)} + \mathfrak{u(1)^{3}}$ and $\mathfrak{g}_{P} =\mathfrak{so(4)}$. Solving the cohomology condition through the $\,\,\mZ^{Q}_{\infty} || \mZ^{P}_{-1}$ branch (see table \ref{tablegQgP}), leaves us with a non-geometric type B configuration (see table \ref{tablePhisystem}). The $\mathrm{ker}(\tilde{\Phi}_{Q})$ is expanded by $(\epsilon_0,\epsilon_3)$ while that of $\tilde{\Phi}(\mathfrak{g}_{P})$ is expanded by $(\rho_0,\rho_3)$ for this pairing. In this case, the NS-NS and R-R fluxes account for six degrees of freedom and generate a flux-induced $C'_{8}$ tadpole given by (\ref{TadI7Delta}) with $\Delta_{Q}=\epsilon_2/3$ and $\Delta_{P}=(\rho_2-\rho_1)/3$.
\\

\begin{table}[htb]
\small{
\renewcommand{\arraystretch}{1.15}
\begin{center}
\begin{footnotesize}
\begin{tabular}{c|c|c|c|c|c}
  & \multicolumn{5}{c}{$\mathfrak{g}_{P}$ deformation} \\
\\

$\mathfrak{g}_{Q}$ original & $\mathfrak{so(4)}$  & $\mathfrak{so(3,1)}$  &     $\mathfrak{su(2)+u(1)^3}$ &    $\mathfrak{iso(3)}$ & $\mathfrak{nil}$  \\

& & & & & \\

\hline

   &      &  &       &  & \\
$\mathfrak{so(4)}$  & $\mZ^{Q}_{-1} \parallel \mZ^{P}_{-1}$   &   $\mZ^{Q}_{-1} \parallel \mZ^{P}_{0}$  & $\mZ^{Q}_{-1} \parallel \mZ^{P}_{\infty}$ &  $\mZ^{Q}_{-1} \parallel \mZ^{P}_{+1}$ &   $\mZ^{Q}_{-1} \parallel \mZ^{P}_{\infty}$ \\
     &       &  &          & &  \\
\hline
 & & &  &  & \\  
$\mathfrak{so(3,1)}$ &  $\mZ^{Q}_{0} \parallel \mZ^{P}_{-1}$ & $\mZ^{Q}_{0} \parallel \mZ^{P}_{0}$   & $\mZ^{Q}_{0} \parallel \mZ^{P}_{\infty}$ & $\mZ^{Q}_{0} \parallel \mZ^{P}_{+1}$& $\mZ^{Q}_{0} \parallel \mZ^{P}_{\infty}$   \\
 & &  &  &  & \\  

\hline

 & & &  & & \\  
$\mathfrak{su(2)+u(1)^3}$&$\mZ^{Q}_{\infty} \parallel \mZ^{P}_{-1} $ & $\mZ^{Q}_{\infty} \parallel \mZ^{P}_{0}$ & $\mZ^{Q}_{\infty} \parallel \mZ^{P}_{\infty} $ & $\mZ^{Q}_{\infty} \parallel \mZ^{P}_{+1}$ &  $\mZ^{Q}_{\infty} \parallel \mZ^{P}_{\infty}$ \\
 &  & &  &  & \\

\hline

 & &  &  & & \\  
$\mathfrak{iso(3)}$ &  $\mZ^{Q}_{+1} \parallel \mZ^{P}_{-1} $ & $\mZ^{Q}_{+1} \parallel \mZ^{P}_{0}$ &   $\mZ^{Q}_{+1} \parallel \mZ^{P}_{\infty} $ & $\mZ^{Q}_{+1} \parallel \mZ^{P}_{+1} $ & $\mZ^{Q}_{+1} \parallel \mZ^{P}_{\infty} $\\
 & &  &  & &  \\

\hline

 & &  &  & & \\  
$\mathfrak{nil}$ &  $\mZ^{Q}_{\infty} \parallel \mZ^{P}_{-1} $ & $\mZ^{Q}_{\infty} \parallel \mZ^{P}_{0}$ &   $\mZ^{Q}_{\infty} \parallel \mZ^{P}_{\infty} $ & $\mZ^{Q}_{\infty} \parallel \mZ^{P}_{+1} $ &  $\mZ^{Q}_{\infty} \parallel \mZ^{P}_{\infty} $\\
 & &  &  &  & \\  

\end{tabular}
\end{footnotesize}
\end{center}
\caption{Roots alignment in non-geometric type B configurations.}
\label{tablePhisystem}
}
\end{table}

The $\bar{H}_{3}$ and $\bar{F}_{3}$ background fluxes determine the flux-induced $\mathcal{P}_2(\mZ_{Q})$ and $\mathcal{P}_{1}(\mZ_{P})$ polynomials in the superpotential. Fixing a non-geometric type A configuration, $\mathcal{P}_2(\mZ_{Q})$ is shown in table \ref{tableP2roots} for each $\mathfrak{g}_Q$ algebra. The equivalent expression for the polynomial $\mathcal{P}_{1}(\mZ_{P})$, resulting from the $\mathfrak{g}_{P}$ algebra, is obtained upon replacing $\epsilon_i \leftrightarrow \rho_i$ and $\mZ_Q \leftrightarrow \mZ_P$.

\begin{table}[htb]
\small{
\renewcommand{\arraystretch}{1.15}
\begin{center}
\begin{tabular}{|c|c|c|c|}
\hline
$\mathfrak{g}_{Q}$ & $\mathcal{P}_2(\mZ_{Q}) \equiv \frac{P_2(U)}{(\gamma_q \, U + \delta_q)^3}$ \\
\hline
\hline
$\mathfrak{so(4)}$ & $\epsilon_3 \, \mZ_{Q}^3 + \epsilon_0$ \\
\hline
$\mathfrak{so(3,1)}$ & $\epsilon_3 \, \mZ_{Q}^3 -3 \, \epsilon_0 \, \mZ_{Q}^2 - 3 \, \epsilon_3 \, \mZ_{Q} + \epsilon_0$  \\
\hline
$\mathfrak{su(2)+u(1)^3}$ & $\epsilon_3 \, \mZ_{Q}^3 + \epsilon_0$ \\
\hline
$\mathfrak{iso(3)}$ &  $\epsilon_1 \, \mZ_{Q} + \epsilon_0$  \\
\hline
$\mathfrak{nil}$ &  $\epsilon_1 \, \mZ_{Q}  + \epsilon_0 $\\
\hline
\end{tabular}
\end{center}
\caption{NS-NS flux-induced polynomials in the non-geometric type A configurations.}
\label{tableP2roots}
}
\end{table}

\section{Supersymmetric solutions.}

In this section, we provide some examples of supersymmetric vacua of the T and S-duality invariant effective supergravity given by the standard K\"ahler potential and the moduli potential induced by R-R $\bar{F}_{3}$, NS-NS $\bar{H}_{3}$ and non-geometric $Q$ and $P$ fluxes using the methods we have developed in this work. We will focus on solutions with the axiodilaton $S$ and K\"ahler $T$ moduli being completely stabilized.
\\

The starting point is the 4d effective theory defined by the K\"ahler potential
\begin{eqnarray}
K &= & -3\,\ln \Big( -i(T-\bar{T}) \Big) -\ln \Big( -i(S-\bar{S}) \Big) -3\,\ln \Big( -i(U-\bar{U}) \Big) 
\end{eqnarray}
and the superpotential (\ref{Wiso}), which can be rewritten as
\begin{eqnarray}
\label{WisoZ}
W &=& -(\gamma_{p} \, U + \delta_{p})^3 \left[  \Big( \sum_{i=0}^{3} \rho_{i} \, \mZ_{P}^{i} \Big ) + 3\,T \,S \,\mathcal{P}_{4}(\mZ_{P}) \right]+ \nonumber \\ &+& (\gamma_{q} \, U + \delta_{q})^3 \left[  S \,  \Big( \sum_{i=0}^{3} \epsilon_{i} \, \mZ_{Q}^{i} \Big ) + 3 \,T \, \mathcal{P}_{3}(\mZ_{Q}) \right] \ ,
\end{eqnarray}
with $\mathcal{P}_{3}(\mZ_{Q})$, $\mathcal{P}_{4}(\mZ_{P})$ taken from table \ref{tableP3roots} according with a fixed pairing $(\mathfrak{g}_{Q},\mathfrak{g}_{P})$ and $\mZ_{Q}$ and $\mZ_{P}$ the modular variables from (\ref{modularZQZP}). In general, $\mZ_{Q} \neq \mZ_{P}$, and we will have to deal with two modular variables instead of just one, $\mZ$. Each pairing $(\mathfrak{g}_Q,\mathfrak{g}_P)$ gives rise to a specific superpotential due to the relationship between the roots structure of a polynomial and its associated algebra.
\\

A supersymmetric vacuum implies the vanishing of the F-terms
\beqa
\label{FFlat}
F_{T} &=& \partial_T W + \frac{3 i W}{2 \,\textrm{Im} T} = 0  \ ,\nonumber \\
F_{S} &=& \partial_S W + \frac{ i W}{2 \,\textrm{Im} S} = 0   \ ,\\
F_{U} &=& \partial_U W + \frac{3 i W}{2 \,\textrm{Im} U} = 0  \ , \nonumber 
\eeqa
bringing about either Minkowski or AdS$_4$ solutions because the potential (\ref{VPotential}) at the minimum is given by $V_0 = - 3 e^{K_0} |W_0|^2 \leq 0$. Restricting our search to Minkowski solutions, i.e. $V_{0}=0$, simplifies (\ref{FFlat}) to
\beq
\label{Minkowskivacuum}
\partial_{S} W = \partial_{T} W=\partial_{U} W = W =0 \ .
\eeq

Working with the generic expression (\ref{Wiso}) for the superpotential, the K\"ahler moduli and axiodilaton equations of motion fix both moduli to
\beqa
\label{STstabilization}
S_{0} &=& -\frac{P_{3}(U_0)}{P_{4}(U_0)} =  \left(\frac{\gamma_{q} \, U + \delta_{q}}{\gamma_{p} \, U + \delta_{p}}\right)^{3}  \,  \left. \frac{\mathcal{P}_{3}(\mZ_{Q})}{ \mathcal{P}_{4}(\mZ_{P}) } \right|_{U_{0}} \nonumber \ ,\\[0mm]
\\
T_{0} &=& -\frac{P_{2}(U_0)}{P_{4}(U_0)}  =  \left(\frac{\gamma_{q} \, U + \delta_{q}}{\gamma_{p} \, U + \delta_{p}}\right)^{3}  \,  \left. \frac{ \sum_{i=0}^{3} \epsilon_{i} \, \mZ_{Q}^{i}  }{ \mathcal{P}_{4}(\mZ_{P}) } \right|_{U_{0}} \ ,  \nonumber
\eeqa
where $S_{0}$, $T_{0}$ and $U_{0}$ are moduli values at the vacuum. These values are subject to physical considerations. Im$S_{0}$ must be positive because it is the inverse of the string coupling constant $g_{s}$. Im$T_{0}=e^{-\phi} A$ where $A$ is the area of a 4-dimensional subtorus, so it also has to be positive. Also, for the modular variables $\mZ_{Q}$ and $\mZ_{P}$ at the minimum, it happens that Im$\mZ_{Q}=$Im$U_{0}|\Gamma_{Q}|/|\gamma_{q} U_{0} + \delta_{q}|^{2}\,$ and Im$\mZ_{P}=$Im$U_{0}|\Gamma_{P}|/|\gamma_{p} U_{0} + \delta_{p}|^{2}\,$. Therefore, necessarily Im$\mZ_{Q}\neq 0$ and Im$\mZ_{P} \not= 0$ because for Im$U_{0}=0$ the internal space is degenerate. Without loss of generality, we choose Im$U_{0}>0$. 
\\

Finally, an interesting question is whether the VEVs for the moduli give rise to an effective supergravity that is a reliable approximation to string theory. In order to exclude non-perturbative string effects, the string coupling constant $g_{s}=1/$Im$S_{0}$ is expected to be small. However, the conventionally expected large internal volume $V_{int}=($Im$T_{0}/$Im$S_{0}$)$^{3/2}$ required to neglect corrections in $\alpha'$ becomes a more delicate issue in the presence of non-geometric fluxes. For instance, in a non-geometric solution the internal space might be a T-fold \cite{Hull1,Hull2} and therefore a large internal volume could imply small cycles, related by T-duality, with light winding modes associated. These new effects are still not well understood and therefore a large internal volume is physically motivated just for geometric solutions (or those that can be described geometrically in some duality frame). In this work we will limit ourselves to searching supersymmetric solutions of an effective field theory without speculating on their lifting to solutions of the full string theory.
\\

The remaining $\,W=0\,$ and $\,\partial_{U} W=0\,$ conditions can be rewritten, using (\ref{STstabilization}), as
\beqa
E(U_{0})=P_1(U_{0}) \, P_4(U_{0}) - P_2(U_{0}) \, P_3(U_{0}) =0 \ ,\\
E'(U_{0})=0 \ ,
\eeqa
provided\footnote{This has to be the case for Im$U_{0}$ $\neq 0$ in all $\mathfrak{g}_P$ but $\mathfrak{g}_P=\mathfrak{so(3,1)}$ that has complex roots $\mZ_{P}=\pm i$. For this singular case, $P_4(U_{0}) = 0$ implies $P_i(U_{0}) = 0$ for $i=1,2,3,4$ as can be seen from (\ref{Minkowskivacuum}). Then $S$ and $T$ cannot be simultaneously stabilized in a supersymmetric Minkowski vacuum.} $P_4(U_{0}) \neq 0$. The prime denotes differentiation with respect to $U$ and, therefore, $E(U)$ has a double root. The root must, given our definition for the K\"{a}hler potential, be complex and therefore $E(U)$ contains a double copy of complex conjugate pairs, accounting for 4 of its 6 roots. Therefore, we have the following factorisation property of $E(U)$,
\beq
\label{Efactorization}
E(U) = (f_{2} \, U^{2}+f_{1} \, U+f_{0}) \,\tilde{E}(U) \ , \\
\eeq
with $\tilde{E}(U) \equiv (g_{2} \, U^{2} + g_{1} \, U + g_{0})^{2}$ accounting for the double root that becomes complex iff $g_{1}^{2}-4\, g_{2}\, g_{0} < 0$.
\\

Information about the nature of the six roots of $E(U)$ can be immediately obtained from the generic superpotential polynomials once a $(\mathfrak{g}_{Q},\mathfrak{g}_{P})$ pairing is chosen and the full set of Bianchi identities, ie. integrability, cohomology and singlet Bianchi constraints, are applied. Four cases are automatically discarded because their $E(U)$ possesses at least four real roots, so they can never have a double complex root for the Minkowski vacua to be physically viable, i.e. Im$U_{0} \neq 0$. The number of real roots for each $(\mathfrak{g}_{Q},\mathfrak{g}_{P})$ pairing is summarized\footnote{Entries in table \ref{tableRootsE} are in one to one correspondence with entries in table \ref{tablegQgP}.} in table \ref{tableRootsE}. A priori, all branches with $E(U)$ having a number of real roots less than three could accommodate supersymmetric Minkowski solutions. This is a necessary but not sufficient condition for the existence of Minkowski vacua because for $E(U)$ to split into the form (\ref{Efactorization}), additional constraints on $\bar{H}_{3}$ and $\bar{F}_{3}$ fluxes are needed. Therefore, several branches in table \ref{tableRootsE} will exclude Minkowski vacua, even though they have a sufficient number of complex roots and we will provide an example of this.
\\

Despite this, several results can be read from table \ref{tableRootsE} : $i)$ There are no supersymmetric Minkowski solutions in the $(\mathfrak{nil},\mathfrak{nil})$ case because all $E(U)$ roots become real for this pairing. $ii)$ For supersymmetric Minkowski solutions to exist in $(\mathfrak{iso(3)},\mathfrak{iso(3)})$, $(\mathfrak{iso(3)},\mathfrak{nil})$ and $(\mathfrak{nil},\mathfrak{iso(3)})$ pairings, it is necessary to have non-geometric type B configurations (see table \ref{tablePhisystem}), generating an eventually non vanishing flux-induced $C'_8 $ tadpole. $iii)$ The rest of the pairings are richer and supersymmetric Minkowski solutions could, in principle, exist in all branches that solve the cohomology condition (see table \ref{tablegQgP}).

\begin{table}[htb]
\small{
\renewcommand{\arraystretch}{1.15}
\begin{center}
\begin{normalsize}
\begin{tabular}{c|c|c|c|c|c}
  & \multicolumn{5}{c}{$\mathfrak{g}_{P}$ deformation} \\

\\

$\mathfrak{g}_{Q}$ original & $\mathfrak{so(4)}$  & $\mathfrak{so(3,1)}$  &     $\mathfrak{su(2)+u(1)^3}$ &    $\mathfrak{iso(3)}$ & $\mathfrak{nil}$  \\

& & & & & \\

\hline

   &    $2$     &  &        $2$&  & \\
$\mathfrak{so(4)}$  &   $2$ &   $1$  &   $2$ &  $1$ &   $1$ \\
     &    $1$   &  &        $1$  & &  \\
\hline
 & & $2$ &  &  & \\  
$\mathfrak{so(3,1)}$ &  $1$ &   $2$   & $1$ & $1$& $1$   \\
 & & $1$ &  &  & \\  

\hline

 &$2$ & &  $2$ & & \\  
$\mathfrak{su(2)+u(1)^3}$& $2$ & $1$ &  & $1$ &  $2$ \\
 & $1$ & & $2$ &  & \\

\hline

 & &  &  & $\textbf{4}$ & $\textbf{4}$ \\  
$\mathfrak{iso(3)}$ &  $1$ & $1$ &   $1$ &  & \\
 & &  &  & $1$ & $1$ \\

\hline

 & &  &  & $\textbf{4}$ & \\  
$\mathfrak{nil}$ &  $1$ & $1$ &   $2$ &  &  $\textbf{6}$\\
 & &  &  & $1$ & \\  

\end{tabular}
\end{normalsize}
\end{center}
\caption{Number of real roots of $E(U)$ defined in ($7.8$) after imposing the full set of Bianchi constraints.}
\label{tableRootsE}
}
\end{table}

\subsection{Our example: $\mathfrak{g}_{Q}=\mathfrak{su(2)+u(1)^{3}}$ deformed by $\mathfrak{g}_{P}=\mathfrak{so(4)}$.}

For our first example, we shall continue to investigate the case $\mathfrak{g}_{Q}=\mathfrak{su(2)+u(1)^{3}}$ deformed by $\mathfrak{g}_{P}=\mathfrak{so(4)}$, in order to show how simple supersymmetric solutions can be easily obtained using these methods.
\\

For the sake of simplicity, we will look for $\bar{H}_{3}$ and $\bar{F}_{3}$ background fluxes with $\vec{\epsilon} \in \mathrm{ker}(\tilde{\Phi}_{Q})$ and $\vec{\rho} \in \mathrm{ker}(\tilde{\Phi}_{P})$, so $N'_{7}=0$ but the net charges $N_{7}$ and $\tilde{N}_{7}$ are considered as free variables. In these solutions, $\mathcal{P}_{2}(\mZ_{Q})$ and $\mathcal{P}_{1}(\mZ_{P})$ can be obtained from table \ref{tableP2roots} leaving us with a set $(\epsilon_0,\epsilon_3 \,;\, \rho_0,\rho_3)$ of free parameters in the superpotential determining the $\bar{H}_{3}$ and $\bar{F}_{3}$ background fluxes.
\\

Taking the relevant polynomials from tables \ref{tableP3roots} and \ref{tableP2roots}, the superpotential (\ref{WisoZ}) becomes
\beqa
\label{Wexample}
W&=& - (\gamma_{p} \, U + \delta_{p})^3 \left[ (\rho_3 \, \mZ_{P}^3 + \rho_0)  + 3\,T \,S \,  \mZ_{P}(\mZ_{P}+1) \right] + \nonumber \\
&+& (\gamma_{q} \, U + \delta_{q})^3  \left[  S \, (\epsilon_3 \, \mZ_Q^3 + \epsilon_0)  + 3\,T\, \mZ_{Q} \right]  
\eeqa
and the tadpole cancellation conditions can be expressed in terms of the roots as
\beqa
\label{tadexample}
N_{3}&=& A_{33} \, (\mZ^{Q}_{0} \times \mZ^{P}_{0})^3 + A_{30}\, (\mZ^{Q}_{\infty} \times \mZ^{P}_{0})^3 + A_{03}\, (\mZ^{Q}_{0} \times \mZ^{P}_{\infty})^3 + A_{00}\, (\mZ^{Q}_{\infty} \times \mZ^{P}_{\infty})^3 \ , \nonumber\\ [2.5mm]
N_{7}&=& \rho_{3} \, (\mZ^{Q}_{0} \times \mZ^{P}_{0}) \,  (\mZ^{Q}_{\infty} \times \mZ^{P}_{0})^2 + \rho_{0} \, (\mZ^{Q}_{0} \times \mZ^{P}_{\infty}) \,  (\mZ^{Q}_{\infty} \times \mZ^{P}_{\infty})^2  \ , \\[2.5mm]
\tilde{N}_{7}&=& -\epsilon_{3} \, (\mZ^{Q}_{0} \times \mZ^{P}_{0}) \,  (\mZ^{Q}_{0} \times \mZ^{P}_{\infty}) \,(\mZ^{Q}_{0} \times \mZ^{P}_{-1}) - \epsilon_{0} \, (\mZ^{Q}_{\infty} \times \mZ^{P}_{0}) \,  (\mZ^{Q}_{\infty} \times \mZ^{P}_{\infty}) \, (\mZ^{Q}_{\infty} \times \mZ^{P}_{-1}) \ ,\nonumber 
\eeqa
with $A_{ij}=-\rho_{i}\,\epsilon_{j}$. We now impose the constraints from one of the prime ideals of the cohomology condition, of which there are three to choose for this pairing, as shown in table \ref{tablegQgP} and explicitly stated in (\ref{2cocycleexample}). The case $J_{1}=0$ is automatically fulfilled with an embedding $\Gamma_{P}=\Gamma_{Q} \equiv \Gamma$, or equivalently $\mZ_{P}=\mZ_{Q} \equiv \mZ$, while the $J_{2}=0$ results are equivalent to this after applying a T-duality transformation $\mZ \rightarrow -1/\mZ$. The case $J_{3}=0$ is a little bit different from the previous ones. It cannot be transformed into $J_{1,2}=0$ and the resultant solutions are distinct from those of the first two branches. We will solve for each of the three branches and clarify their relation to the existence of both AdS$_{4}$ and Minkowski vacua.

\subsubsection{Simple type A AdS$_{4}$ solutions.}

Imposing\footnote{Imposing $J_{2}=0$ is T-dual to $J_{1}=0$ just with $\mZ \rightarrow -1/\mZ $.} $\,J_{1}=0\,$ and just fixing the modular embeddings to be 
\begin{equation}
\label{GammaPeqGammaQ}
\Gamma_{P} = \Gamma_{Q} \equiv \Gamma=
\left(
       \begin{array}{cc}
            \alpha  &  \beta \\
            \gamma  & \delta   
       \end{array}
\right) \ ,
\end{equation} 
provides us with a much simplified superpotential, given by
\beq
\label{Wexamplebranch1}
\frac{W}{(\gamma \, U + \delta)^3} = -(\rho_3 \, \mZ^3 + \rho_0)  + S \, (\epsilon_3 \, \mZ^3 + \epsilon_0) + 3\,T\, \mZ - 3\,T \,S \,\mZ \, (\mZ+1) \ .
\eeq

Under the transformation $U \rightarrow \mZ$, we have $e^{K}W \rightarrow e^{\mathcal{K}}\mathcal{W}$ with
\beqa
\mathcal{K} &=& -3\,\ln \Big( -i(T-\bar{T}) \Big) -\ln \Big( -i(S-\bar{S}) \Big) -3\,\ln \Big( -i(\mZ-\bar{\mZ}) \Big) \ ,\\[2mm]
\mathcal{W} &=& |\Gamma|^{3/2} \left[ - (\rho_3 \, \mZ^3 + \rho_0)  + S \, (\epsilon_3 \, \mZ^3 + \epsilon_0) + 3\,T\, \mZ - 3\,T \,S \,\mZ \, (\mZ+1)  \right]
\eeqa
and the tadpole cancellation conditions (\ref{tadexample}) simplify to
\beqa
\label{tadexamplebranch1}
N_{3}&=& |\Gamma|^{3} (A_{03}-A_{30}) =|\Gamma|^{3} (\epsilon_0\,\rho_3-\epsilon_3\,\rho_0) \ ,\\
N_{7}&=& \tilde{N}_{7} = 0 \ .
\eeqa

It is worth noting that, by simply imposing the embedding (\ref{GammaPeqGammaQ}), it becomes impossible to have non-geometric type B configurations, as we can see from table \ref{tablePhisystem}. Indeed, the alignment $ \, \mZ_{\infty} || \mZ_{-1}$ results in $|\Gamma|=0 \, $ and the isomorphism is no longer valid. So whenever we impose (\ref{GammaPeqGammaQ}), automatically $\epsilon_1=\epsilon_2=\rho_1=\rho_2=0$ and then $N'_{7}=N_{7}= \tilde{N}_{7} = 0$.
\\

It can also be proven that this system does not possess Minkowski vacua. To do this, let us compute restrictions on the NS-NS $\bar{H}_{3}$ and R-R $\bar{F}_{3}$ background fluxes needed for the polynomial $E(U)$ to be factorized as (\ref{Efactorization}). From table \ref{tableRootsE} we know that $E(U)$ has at least two real roots. Factorising  out and dropping these real roots, $E(U)\rightarrow \tilde{E}(U)$, it can be shown that for $\tilde{E}(U)$ to possess a double complex root, the $\bar{H}_{3}$ and $\bar{F}_{3}$ background fluxes must satisfy
\beqa
8 \, \epsilon_0 \, \rho_3 + (\epsilon_3 - 9 \, \rho_3) \, \rho_0 &=& 0 \ , \\
(\epsilon_3 - \rho_3)^{3} - 8 \, \rho_3^{2} \, \rho_0 &=& 0 \ ,
\eeqa
and so $g_{1}^{2}-4\, g_{2}\, g_{0} = 12 \,\Big(\rho_3^{1/3} \,\rho_0^{1/3}\Big)^{2} \geq 0$, fixing all six roots of $E(U)$ to be real and producing non physical vacua, i.e. Im$U_{0}=0$.
\\

However, we find that supersymmetric AdS$_4$ vacua can exist without introducing localized sources. This result is new compared to the T-duality invariant effective theory which was deeply studied in \cite{Font:2008vd}. Let us fix $\epsilon_3=\rho_3=0$ so as to have $N_{3}=0$ and, for instance, $\rho_0=2 \, \epsilon_0$. Solving the F-flat conditions (\ref{FFlat}) we obtain
\beq
\mZ_{0}= -1.0434 +  0.4758 \,i \hspace{3mm},\hspace{3mm} S_{0}=-2.3802 +  4.1685 \,i \hspace{3mm},\hspace{3mm} \epsilon_{0}^{-1}\,T_{0}=-0.4022 + 1.1483 \,i \ ,
\eeq
with a vacuum energy $V_{0}\,\epsilon_{0}/|\Gamma|^{3}=-2.3958$ and with $N_{3}=N_{7}=\tilde{N}_{7}=N'_{7}=0$. In terms of the original complex structure modulus, $\,U_{0}= \Gamma^{-1}\,\mZ_{0}\,$ with $\Gamma$ the modular matrix given in (\ref{GammaPeqGammaQ}). Fixing for example $\beta=\gamma=0$, this solution corresponds to $a_{0}= -2 \, \epsilon_{0} \, \delta^{3}$, $b_{0}= - \epsilon_{0} \, \delta^{3}$, $\tilde{c}_{1}=\tilde{d}_{1}=-\alpha\,\delta^{2}$ and $\tilde{d}_{2}=\alpha^2 \, \delta$. Large positive values of the $\epsilon_0$ parameter translate into large absolute values of the NS-NS and R-R fluxes and also a large internal volume.

\subsubsection{Simple type B Minkowski solutions.}

Now we explore the case $J_{3}=0$, or equivalently $\mZ^{Q}_{\infty} || \mZ^{P}_{-1}$. As an example of this alignment involving just two modular parameters let us take
\begin{equation}
\Gamma_{Q} =
\left(
       \begin{array}{cc}
            \alpha  &  -\delta \\
            \alpha  & \delta   
       \end{array}
\right)
\hspace{5mm},\hspace{5mm}
\Gamma_P =
\left(
       \begin{array}{cc}
            \alpha  &  0 \\
            0  &\delta   
       \end{array}
\right) \ .
\end{equation} 

This results in a two dimensional family of non-geometric type B configurations. Substituting directly in (\ref{STstabilization}) we obtain
\beq
\label{linearfieldsexample}
T_0=\frac{1}{3 \, \alpha \,  \delta} \, \frac{\epsilon_3 \, (\alpha  U_0 -\delta )^3+\epsilon_0 \, (\delta +\alpha  U_0 )^3}{  U_0 \,  (\delta +\alpha  U_0 )} \hspace{5mm},\hspace{5mm} S_0= \frac{\alpha \, U_0}{\delta }-\frac{\delta }{\alpha  U_0 } \ .
\eeq

Let us compute again restrictions on the NS-NS $\bar{H}_{3}$ and R-R $\bar{F}_{3}$ background fluxes needed for the polynomial $E(U)$ to be factorized as (\ref{Efactorization}). From table \ref{tableRootsE}, this time $E(U)$ has at least one real root. Factorising out this real root,  $E(U) \rightarrow (f_1\,U + f_0)\,\tilde{E}(U)$, this imposes
\beq
\rho_0=0 \hspace{3mm},\hspace{3mm} \epsilon_0 = -\epsilon_3 = \frac{\rho_{3}}{8} \hspace{3mm},\hspace{3mm} f_{1}=g_{1}=0 \hspace{3mm},\hspace{3mm}\frac{g_0}{g_2}= \left( \frac{\delta}{\alpha}\right)^{2}
\eeq
and therefore $g_{1}^{2}-4\, g_{2}\, g_{0} < 0$, producing physical vacua $U_{0}= i \left( \frac{\delta}{\alpha}\right) $. Substituting directly in (\ref{linearfieldsexample}), the moduli get stabilized to
\beqa
U_0=\left( \frac{\delta}{\alpha} \right)  \,  i \hspace{6 mm},\hspace{6 mm}  S_0= 2 \, i  \hspace{6 mm},\hspace{6 mm}   T_0= \frac{\rho_3}{12} \,(1 +  i) \ .
\eeqa

This family is physical for $\rho_{3} > 0$ and $|\Gamma_{P}|>0$. The tadpole conditions for these supersymmetric Minkowski vacua are
\beq
N_{3}=\frac{\rho_{3}}{4} \hspace{3mm},\hspace{3mm} N_{7}=\rho_{3} \hspace{3mm},\hspace{3mm} \tilde{N}_{7}=|\Gamma_{P}|^{3}\,\frac{\rho_{3}^{2}}{4} \ ,
\eeq
with $|\Gamma_{P}|=\alpha\, \delta$, so $N_3>0$, $N_7>0$ and $\tilde{N}_7>0$ is required.
\\

In terms of the original fluxes, this solution corresponds to $c_{3}=-\alpha^{3}$, $c_{2}=\tilde{c}_{2}=-\tilde{d}_{2}=-\alpha^{2}\, \delta$, $c_{1}=\tilde{c}_{1}=\tilde{d}_{1}=-\alpha\, \delta^{2}$ and $c_{0}=-\delta^{3}$ for non-geometric fluxes; $b_{0}=-\delta^{3}\,\frac{\rho_{3}}{4}$ and $b_{2}=-\alpha^{2}\,\delta\,\frac{\rho_{3}}{4}$ for the NS-NS flux; and $a_{3}=\alpha^{3}\,\rho_{3}$ for the R-R flux. Again, large values of the $\rho_3$ parameter translate into large absolute values of the NS-NS and R-R fluxes and a large internal volume. However, this also increases the number of localized sources and therefore their backreaction, which we are not taking into account.

\subsection{More type B Minkowski vacua examples.}

In our previous example, we gave simple Minkowski solutions with all moduli stabilized in a physical vacuum with a vanishing flux-induced $C'_8$ tadpole, ie. $N'_7=0$. Now, we provide Minkowski solutions with $N'_7 \neq 0$ (examples 3 and 4).
\\

Our main goal in this work has been to develop a systematic method to compute supersymmetric Minkowski vacua based on different $(\mathfrak{g}_{Q},\mathfrak{g}_{P})$ pairings which fulfil all algebraic constraints. To show how these methods work, we conclude by presenting several simple non-geometric type B configurations involving all the six dimensional Lie algebras compatible with the orbifold symmetries. Besides finding analytic VEVs for the moduli, we also relate them to the net charge of localized sources which can exist, as well as some features of such vacua. 

\subsubsection{Example 1: vacua with unstabilized complex structure modulus.}

Let us work out a simple family of Minkowski solutions with a vanishing flux-induced $C'_8$ tadpole for which all the moduli but the complex structure modulus are fixed by the fluxes. These solutions were previously found in \cite{Chen:2007af} and we now clarify their flux structure.
\\

Let us fix the non-geometric $Q$ and $P$ fluxes to be isomorphic to $\mathfrak{g}_{Q}=\mathfrak{so(4)}$ and $\mathfrak{g}_{P}=\mathfrak{iso(3)}$ respectively\footnote{In this case, $\Delta_{Q}=(\epsilon_2-\epsilon_1)/3$ and $\Delta_{P} = -\rho_3-\rho_2/3$.} under the modular embeddings
\beqa
\label{Gammaexample3}
\Gamma_{Q} =
\left(
       \begin{array}{cc}
            \alpha_{q}  &  0 \\
             0  & \delta_{q} 
       \end{array}
\right)
\hspace{5mm},\hspace{5mm}
\Gamma_P =
\left(
       \begin{array}{cc}
            \alpha_{p}  &  \beta_{p} \\
            0  &\delta_{p}
       \end{array}
\right) \ .
\eeqa

The cohomology condition for this pairing has an unique branch (see table \ref{tablegQgP}) implying $\,\mZ^{Q}_{-1} \parallel \mZ^{P}_{+1}\,$ and it is a type B configuration, as is shown in table \ref{tablePhisystem}. This relates the modular matrices (\ref{Gammaexample3}) so that, $\,\alpha_{q}= \lambda \, \alpha_{p}\,$ and $\,\delta_{q}=\lambda\,(\beta_{p}-\delta_{p})\,$.
\\

Taking for simplicity $\vec{\epsilon} \in \mathrm{ker}(\tilde{\Phi}_{Q})$ and $\vec{\rho} \in \mathrm{ker}(\tilde{\Phi}_{P})$ results in $\epsilon_{1}=\epsilon_{2}=\rho_{2}=\rho_{3}=0$. Moreover, we will also fix $\epsilon_{3}=0$ and therefore, substituting into (\ref{STstabilization}), we obtain
\beq
\label{STexample3}
S_{0}=-\lambda^{3} \left( \frac{\alpha_{p}}{\delta_{p}^{2}}\right)\,(\beta_{p}-\delta_{p}) \, U_{0} \hspace{3mm},\hspace{3mm} T_{0}=-\frac{\lambda^{3}\,\epsilon_0\,(\beta_{p}-\delta_{p})^{3}}{3\, \delta_{p}^{2}\,(\alpha_{p}\,U_{0}+(\beta_{p}-\delta_{p}))} \ .
\eeq

Upon substituting these axiodilaton and K\"ahler moduli VEVs into the superpotential we have
\beq
\label{WUexample3}
W(U_{0})=- \left( \frac{\alpha_{p}}{\delta_{p}^{2}}\right)\,\Big( \lambda^{6}\,(\beta_{p}-\delta_{p})^{4}\, \epsilon_0 + \delta_{p}^{4}\, \rho_1 \Big)\,U_{0} - \delta_{p}^{2} \, \Big( \delta_{p}\, \rho_0 + \beta_{p}\, \rho_1  \Big)  .
\eeq

For Minkowski solutions to exist $\,\partial_{U} W = W = 0\,$. Moreover, because of $\alpha_{p}\,\delta_{p} \neq 0$ (otherwise $|\Gamma_{P}|=0$), Minkowski vacua with complex structure modulus unstabilized do exist provided
\beqa
\lambda^{6}\,(\beta_{p}-\delta_{p})^{4}\, \epsilon_0 + \delta_{p}^{4}\, \rho_1 &=& 0 \ , \\
\delta_{p}\, \rho_0 + \beta_{p}\, \rho_1 &=& 0 \ .
\eeqa 

Under these restrictions for $\rho_0$ and $\rho_1$, the tadpole cancellation conditions simplify to
\beqa
N_{3}&=&\tilde{N}_{7}=N'_{7}=0 \ ,\\
N_{7}&=&\frac{\lambda^{9}\,\epsilon_0}{3}\,\left( \frac{\alpha_{p}^{3}}{\delta_{p}^{2}}\right)\,(\beta_{p}-\delta_{p})^{5} \ . 
\eeqa

From the axiodilaton and K\"ahler stabilization (\ref{STexample3}), taking a physical vacuum with Im$U_{0}>0\,$ implies
\beqa
\lambda\, \alpha_{p} \, (\beta_{p}-\delta_{p}) &<& 0  \ , \\
\lambda\, \alpha_{p} \, (\beta_{p}-\delta_{p}) \,\epsilon_0 &>& 0 \ , 
\eeqa 
for Im$S_{0}>0$ and Im$T_{0}>0$ and therefore $\epsilon_0<0$. Otherwise the vacuum is not physical. Therefore $N_{7}>0$ and so D7-branes are needed. Several configurations of these necessary D7-branes were presented in \cite{Chen:2007af}. Large values of $|\lambda|$ and $|\epsilon_{0}|$ favor a small string coupling constant and increase the internal volume, ie. $g_{s} \propto 1/|\lambda|^{3}$ and $V_{int} \propto |\epsilon_0|^{3/2}$, for a fixed $\Gamma_{P}$ modular matrix and a given VEV for the complex structure modulus, $U_{0}$.

\subsubsection{Example 2: vacua with a hierarchy of fluxes.}

In this example we work out a family of solutions with a richer structure of localized sources. This time we fix the non-geometric $Q$ and $P$ fluxes to be isomorphic to $\mathfrak{g}_{Q}=\mathfrak{so(4)}$ and $\mathfrak{g}_{P}=\mathfrak{so(4)}$ respectively\footnote{In this case $\Delta_{Q}=(\epsilon_2-\epsilon_{1})/3$ and $\Delta_{P} = (\rho_2-\rho_1)/3$.}.
\\

Just to illustrate some vacua with this algebraic structure, we fix the modular embeddings to be
\beqa
\label{Gammaexample5}
\Gamma_{Q} =
\left(
       \begin{array}{cc}
            \alpha  &  \delta \\
             -\lambda \, \alpha  & \lambda \,\delta 
       \end{array}
\right)
\hspace{5mm},\hspace{5mm}
\Gamma_P =
\left(
       \begin{array}{cc}
            (1-\lambda)\,\alpha  &  0 \\
            0  &(1+\lambda)\,\delta
       \end{array}
\right) \ ,
\eeqa
with $\alpha \, \delta\neq 0$ and $\lambda\,(1-\lambda^{2}) \neq 0$ for the isomorphism to be well defined.
\\

The cohomology condition has three branches, as seen in table \ref{tablegQgP}, and the embeddings (\ref{Gammaexample5}) satisfy $\,\mZ^{Q}_{-1} \parallel \mZ^{P}_{-1}\,$, giving a type B configuration (see table \ref{tablePhisystem}). For simplicity we will fix again $\vec{\epsilon} \in \mathrm{ker}(\tilde{\Phi}_{Q})$ and $\vec{\rho} \in \mathrm{ker}(\tilde{\Phi}_{P})$, and so this time $\epsilon_1=\epsilon_2=\rho_1=\rho_2=0$ which results in $N'_{7}=0$. Under this fluxes choice, $E(U)$ has 1 real root, as table \ref{tableRootsE} shows. We find that $E(U)$ can be factorized as (\ref{Efactorization}) with $g_{1}=0$ and 
\beqa
\epsilon_3 &=& \frac{1-\lambda ^2}{8 \lambda } \Big(\,(\lambda -1)^3 \rho _3 +(\lambda +1)^3 \rho _0\,\Big) \ , \\[2mm]
\epsilon_0 &=& \frac{1-\lambda ^2 }{8 \lambda ^4}\Big(\,(\lambda -1)^3 \rho _3-(\lambda +1)^3 \rho _0 \, \Big) \ , \\[2mm]
\frac{g_0}{g_2}&=& \left( \frac{\delta}{\alpha}\right)^{2} \hspace{2mm} , \hspace{2mm} \frac{f_0}{f_1}=-\left( \frac{\delta}{\alpha}\right)\frac{(\lambda -1)^3 \rho _3}{(\lambda +1)^3 \rho _0} \ . 
\eeqa

Since $g_{1}^{2}-4\, g_{2}\, g_{0} < 0$, these are physical vacua with $U_{0}= i \left( \frac{\delta}{\alpha}\right) $. From (\ref{STstabilization}), the axiodilaton and K\"ahler moduli get stabilized to

\beqa
S_0 &=& \left( \frac{2 \lambda}{\lambda^{2}-1}\right) i  \ , \nonumber\\
\\
T_0 &=&\frac{\lambda^2-1}{12 \lambda (\lambda^{2}+1)} \Big(\,\frac{(\lambda +1)^4}{\lambda^2-1} \, \rho _0-\frac{(\lambda -1)^4}{\lambda^2-1} \, \rho _3 \,+\, i  \left(\, (\lambda -1)^2 \, \rho _3 +(\lambda +1)^2 \rho _0\,\right) \, \Big) \ . \nonumber
\eeqa

The resultant tadpole conditions for these vacua are
\beqa
N_{3}&=&\frac{|\Gamma_{Q}|^{3}}{2 \lambda^{2}} (\lambda^{2} - 1) \left(\,   (\lambda -1)^6 \, \tilde{\rho}_{3}^2 + (\lambda +1)^6 \,\tilde{\rho}_{0}^2  \, \right) \ ,\\
N_{7}&=&\frac{|\Gamma_{Q}|^{3}}{2 \lambda} (\lambda^{2} - 1)  \left( \,  (\lambda -1)^2 \, \tilde{\rho}_{3} +  (\lambda+1)^2 \, \tilde{\rho}_{0} \, \right) \ ,\\
\tilde{N}_{7}&=& \frac{|\Gamma_{Q}|^{3}}{8 \lambda^{3}} \left( \lambda ^2-1\right)^3 \left( \,  (\lambda -1)^2 \, \tilde{\rho}_{3} +  (\lambda +1)^2\, \tilde{\rho}_{0} \, \right) \ ,
\eeqa
with $\rho_{3}=4\lambda \tilde{\rho}_{3}$ and $\rho_{0}=4\lambda \tilde{\rho}_{0}$. Then $N_{3}>0$, $N_{7}>0$ and $\tilde{N}_{7}>0$ is necessary for vacua to be physical\footnote{Fixing $|\Gamma_{Q}|>0$ implies $\lambda>0$ for Im$U_{0}>0$, $(\lambda^{2}-1)>0$ for Im$S_{0}>0$ and $(\lambda -1)^2 \, \tilde{\rho}_{3} +  (\lambda +1)^2\, \tilde{\rho}_{0}>0$ for Im$T_{0}>0$. This fixes the net charge of the tadpoles.}.
\\

In terms of the original fluxes, this solution corresponds to $c_{3}=-\alpha^{3} \, \lambda \, (\lambda-1)$, $c_{2}=\tilde{c}_{2}=\alpha^{2} \, \delta \, \lambda \, (\lambda+1)$, $c_{1}=\tilde{c}_{1}= -\alpha \, \delta^{2} \, \lambda \, (\lambda-1)$, $c_{0}=\delta^{3} \, \lambda \, (\lambda+1)$ and $\tilde{d}_{1}=\alpha \, \delta^{2}\,  (\lambda^{2}-1)\, (\lambda+1)$, $\tilde{d}_{2}=\alpha^{2} \, \delta \, (\lambda^{2}-1)\, (\lambda-1)$ for non-geometric fluxes; $b_{0}=\delta^{3}\, (\lambda^{2}-1)\, (\lambda-1)^{3} \, \tilde{\rho}_{3}$, $b_{1}=-\alpha \, \delta^{2}\, (\lambda^{2}-1)\,(\lambda+1)^{3}\, \tilde{\rho}_{0}$, $b_{2}=\alpha^{2}\,\delta\,(\lambda^{2}-1)\,(\lambda-1)^{3} \,\tilde{\rho}_{3}$ and $b_{3}=-\alpha^{3}\,(\lambda^{2}-1)\,(\lambda+1)^{3} \,\tilde{\rho}_{0}$ for NS-NS flux and $a_{0}=- 4 \,\delta^{3} \,\lambda\, (\lambda+1)^{3} \,\tilde{\rho}_{0}$ and $a_{3}=- 4 \, \alpha^{3} \, \lambda \, (\lambda-1)^{3} \, \tilde{\rho}_{3}$ for R-R flux.
\\

By considering the fluxes' dependency on the parameter $\lambda$, we note that generically a hierarchy between $\bar{F}_{3},\bar{H}_{3}$ and non-geometric $Q,P$ fluxes occurs, in which the NS-NS and R-R fluxes, i.e. $a_{i}\propto \lambda^{4}$, $b_{j} \propto \lambda^{5}$ are large compared to the non-geometric fluxes, i.e. $c_{i}\propto \lambda^{2}$, $d_{j} \propto \lambda^{3}$, given $\lambda>1$ for Im$S_{0}>0$. However, there is a critical value $\lambda_{0}=1+\sqrt{2}$ for which $g_{s} \geq 1$ if $\lambda \geq \lambda_{0}$. Hence, there is a narrow range,  $1<\lambda<\lambda_{0}$, for which non-perturbative string effects can be neglected, ie. $\lambda=2$ implies $g_{s}=3/4$. Finally, large values of the $\tilde{\rho}_{0}$ and $\tilde{\rho}_{3}$ parameters favor a large internal volume.

\subsubsection{Example 3: vacua with a non vanishing flux-induced $C'_{8}$ tadpole.}

We now consider a simple family of solutions with a non vanishing flux-induced $C'_8$ tadpole for which all moduli get stabilized. Let us fix the non-geometric $Q$ and $P$ fluxes to be isomorphic to $\mathfrak{g}_{Q}=\mathfrak{so(3,1)}$ and $\mathfrak{g}_{P}=\mathfrak{so(4)}$ respectively\footnote{In this case $\Delta_{Q}=-\epsilon_{2}/3-\epsilon_0=-\epsilon'_0/24$ and $\Delta_{P} = (\rho_2-\rho_1)/3$.}. Examples belonging to this pairing were also found in \cite{Aldazabal:2006up}. 
\\

For simplicity, we fix the modular embeddings to be
\beqa
\label{Gammaexample4}
\Gamma_{Q} =
\left(
       \begin{array}{cc}
            \alpha  &  \delta \\
             \alpha  & -\delta 
       \end{array}
\right)
\hspace{5mm},\hspace{5mm}
\Gamma_P =
\left(
       \begin{array}{cc}
            \alpha  &  0 \\
            0  &\delta
       \end{array}
\right) \ .
\eeqa

The cohomology condition for this pairing has an unique branch $\,\mZ^{Q}_{0} \parallel \mZ^{P}_{-1}\,$ as is shown in table \ref{tablegQgP}. It is a non-geometric type B configuration (see table \ref{tablePhisystem}) and therefore, has a potentially non vanishing flux-induced $C'_{8}$ tadpole. The modular embeddings (\ref{Gammaexample4}) belong to this branch.
\\

First of all, we will redefine our NS-NS flux parameters as
\beq
\left(
\begin{array}{c}
 \epsilon'_{3}  \\
 \epsilon'_{1}  \\
 \epsilon'_{2}  \\
 \epsilon'_{0} 
\end{array}
\right)
=
8\,\left(
\begin{array}{cccc}
 3 & 1 & 0 & 0 \\
 3 & -1 & 0 & 0 \\
 0 & 0 & -1 & 3 \\
 0 & 0 & 1 & 3
\end{array}
\right)
\left(
\begin{array}{c}
 \epsilon_{3}  \\
 \epsilon_{1}  \\
 \epsilon_{2}  \\
 \epsilon_{0} 
\end{array}
\right) \ .
\eeq

Solutions with NS-NS and R-R fluxes for which $\vec{\epsilon} \notin \mathrm{ker}(\tilde{\Phi}_{Q})$ and $\vec{\rho} \notin \mathrm{ker}(\tilde{\Phi}_{P})$ can be given parametrically in terms of the $(\kappa_1 , \kappa_2)$ parameters as
\beq
\epsilon'_{3}=\kappa_1 + \kappa_2 \hspace{3mm},\hspace{3mm} \epsilon'_{0}=\kappa_1 - \kappa_2 \hspace{3mm},\hspace{3mm} \rho_{1}=\kappa_2  \hspace{3mm},\hspace{3mm} \rho_{2}=\kappa_1 \ ,
\eeq
with $(\epsilon'_{1},\epsilon'_{2})$ expanding the $\mathrm{ker}(\tilde{\Phi}_{Q})$ and $(\rho_{0},\rho_{3})$ expanding the $\mathrm{ker}(\tilde{\Phi}_{P})$ being completely free. For simplicity, we will deal just with a non vanishing $\kappa_{2}$ parameter plus the R-R fluxes $\rho_0$ and $\rho_3$. All the Bianchi identities are by construction satisfied. In general, $E(U)$ has 1 real root (see table \ref{tableRootsE}) for this algebra pairing, but under this specific fluxes configuration, it has two real roots. Factorising out these real roots, $E(U) \rightarrow \tilde{E}(U)$, and requiring it to factorise as (\ref{Efactorization}) we find
\beq
f_{1}=g_{1}=\rho_0=0 \hspace{3mm},\hspace{3mm} \rho_3=\frac{4}{3} \kappa_{2} \hspace{3mm},\hspace{3mm}\frac{g_0}{g_2}= \left( \frac{\delta}{\sqrt{2}\alpha}\right)^{2}  \hspace{3mm},\hspace{3mm}    f_{0} \, g_{2}^{2}=-16 \, \alpha ^4 \, \delta \, \kappa_{2} \ .
\eeq
These values give $g_{1}^{2}-4\, g_{2}\, g_{0} < 0$, producing physical vacua with $U_{0}= i \left( \frac{\delta}{\sqrt{2}\alpha}\right) $. 
\\

Using (\ref{STstabilization}), the moduli get stabilized to
\beqa
U_0=(\frac{\delta}{\sqrt{2}\alpha}) \,  i \hspace{6 mm},\hspace{6 mm}  S_0= \sqrt{2} \, i  \hspace{6 mm},\hspace{6 mm}   T_0=- \frac{\kappa_{2}}{27} \,(1 +  \sqrt{2}\, i) \ ,
\eeqa
which is physical for $\kappa_{2} < 0$ and $|\Gamma_{P}|>0$. The tadpole conditions for these vacua are
\beqa
\tilde{N}_{7}&=&0 \ ,\\
N_{3}&=&\frac{\kappa_{2}}{15}\,N_{7}=-\frac{\kappa_{2}}{3}\,N'_{7}=\frac{2}{9}\,|\Gamma_{P}|^{3}\,\kappa_{2}^{2} \ , 
\eeqa
with $|\Gamma_{P}|=\alpha\, \delta$, so $N_3>0$, $N_7<0$ and $N'_7>0$ is required.
\\

In terms of the original fluxes, this solution corresponds to $c_{3}=2 \, \alpha^{3}$, $c_{2}=\tilde{c}_{2}=2\,\tilde{d}_{2}=2\,\alpha^{2}\, \delta$, $c_{1}=\tilde{c}_{1}=2 \,\tilde{d}_{1}=-2 \,\alpha\, \delta^{2}$ and $c_{0}=-2 \,\delta^{3}$ for non-geometric fluxes; $b_{0}=-\frac{\kappa_{2}}{6}\,\delta^{3}$ for NS-NS flux and $a_{3}=\frac{4}{3}\,\kappa_{2}\,\alpha^{3}$, $a_{1}=\frac{1}{3}\,\kappa_{2}\,\alpha\, \delta^{2}$ for R-R flux. The string coupling constant turns out to be $g_{s}=1/\sqrt{2}$ and the internal volume increases for large values of $|\kappa_{2}|$. This also increases the number of localized sources cancelling the flux-induced tadpoles.

\subsubsection{Example 4: vacua with a non defined flux-induced $C_{8}$ tadpole sign.}

Finally, and for the sake of completeness, we fix the non-geometric $Q$ and $P$ fluxes to be isomorphic to $\mathfrak{g}_{Q}=\mathfrak{so(4)}$ and $\mathfrak{g}_{P}=\mathfrak{nil}$ respectively\footnote{In this case $\Delta_{Q}=(\epsilon_2-\epsilon_1)/3$ and $\Delta_{P} = -\rho_3$.}.
\\

Now we fix the modular embeddings to be
\beqa
\label{Gammaexample6}
\Gamma_{Q} =
\left(
       \begin{array}{cc}
            \alpha  & 0 \\
             0  & \delta 
       \end{array}
\right)
\hspace{5mm},\hspace{5mm}
\Gamma_P =
\left(
       \begin{array}{cc}
            \alpha  &  -\delta \\
            \alpha  & \delta
       \end{array}
\right) \ ,
\eeqa
with $\alpha \, \delta\neq 0$ for the isomorphism to be well defined.
\\

In this case, we obtained a single cohomology condition, see table \ref{tablegQgP}, $\,\mZ^{Q}_{-1} \parallel \mZ^{P}_{\infty}\,$ which is satisfied by (\ref{Gammaexample6}) and is again a type B configuration (see table \ref{tablePhisystem}).  Once more, solutions with NS-NS and R-R fluxes for which $\vec{\epsilon} \notin \mathrm{ker}(\tilde{\Phi}_{Q})$ and $\vec{\rho} \notin \mathrm{ker}(\tilde{\Phi}_{P})$ can be given parametrically, 
\beq
\epsilon_{1}=-4 \, ( \kappa_1 - 3  \, \kappa_2) \hspace{3mm},\hspace{3mm} \epsilon_{2}= -4 \, ( \kappa_1 + 3  \, \kappa_2) \hspace{3mm},\hspace{3mm} \rho_{3}=\kappa_2  \hspace{3mm},\hspace{3mm} \rho_{2}=\kappa_1 \ ,
\eeq
depending on the $(\kappa_1 , \kappa_2)$ parameters and with $(\epsilon_{0},\epsilon_{3})$ expanding the $\mathrm{ker}(\tilde{\Phi}_{Q})$ and $(\rho_{0},\rho_{1})$ expanding the $\mathrm{ker}(\tilde{\Phi}_{P})$, being completely free.
\\

For this pairing, $E(U)$ has 1 real root as was shown in table \ref{tableRootsE}. Again, we find that $E(U)$ can be factorized as (\ref{Efactorization}) with $g_{1}=f_{1}=0$ and 
\beqa
\epsilon_3=\frac{2 B^2}{A}-2 B+4 A \hspace{2mm}&,&\hspace{2mm} \epsilon_0=-4 A \ , \\[1.5mm]
\frac{g_0}{g_2}= \left( \frac{\delta}{\alpha}\right)^{2} \frac{A}{B} \hspace{2mm} &,& \hspace{2mm} f_0 g_{0}^{2}=-2 A \delta^{5} \ , \\[1.5mm]
\kappa_{1}= \frac{1}{4} (B-5 A) \hspace{2mm} &,& \hspace{2mm} \kappa_{2}=\frac{B-A}{4}  \ , 
\eeqa
with $A=\rho_1 - \rho_0$ and $B=\rho_1 - 5 \rho_0$. Then $g_{1}^{2}-4\, g_{2}\, g_{0} < 0$ provided $AB>0$ and there are physical vacua with $U_{0}= i \left( \frac{\delta}{\alpha}\right) \left( \frac{\sqrt{A}}{\sqrt{B}} \right)$. From (\ref{STstabilization}), the axiodilaton and K\"ahler moduli get stabilized to

\beqa
S_0 = \frac{\sqrt{A} \sqrt{B} }{(A+B)^{2}} \Big( \,2 \sqrt{A} \sqrt{B} + i\,  (B-A) \, \Big) \hspace{3mm},\hspace{3mm} T_0= \frac{4 A }{3 (A+B)} \Big( A + i \, \sqrt{A} \sqrt{B}   \Big) \ . \\[0mm]
\nonumber
\eeqa

The resultant tadpole conditions for these vacua are
\beqa
N_{3}&=&\frac{16}{3} |\Gamma_{Q}|^{3} \left(\, (B-A)^{2} + A B  \, \right) \ , \\[1mm]
N_{7}&=& -\frac{2}{3} |\Gamma_{Q}|^{3} (B-2 A)  \ ,\\[1mm]
\tilde{N}_{7}&=& -|\Gamma_{Q}|^{3} \frac{2 (A+B)^2 }{A} \ ,\\[1mm]
N'_{7}&=& -4 |\Gamma_{Q}|^{3} (B-A) \ ,
\eeqa
from which it follows that $N_{3}>0$, $N_{7}$ has no defined sign, $\tilde{N}_{7}<0$ and $N'_{7}<0$ is required for physical vacua\footnote{Fixing $|\Gamma_{Q}|>0$, then $A,B>0$ for Im$T_{0}>0$ and $(B-A)>0$ for Im$S_{0}>0$. This fixes the net charge of tadpoles but $N_{7}$ depends on the sign of $(B-2A)$, with $N_{7}>0$ for $(B-2A)<0$ and $N_{7}<0$ for $(B-2A)>0$.}.
\\

In terms of the original fluxes, this solution corresponds to $d_{3}=-\alpha^{3}$, $-d_{2}=\tilde{d}_{2}=\tilde{c}_{2}= \alpha^{2}\, \delta$, $d_{1}=-\tilde{d}_{1}=-\tilde{c}_{1}= \alpha\, \delta^{2}$ and $d_{0}=\delta^{3}$ for non-geometric fluxes; $b_{0}=4 \,\delta ^3 \,A $, $b_{1}=\frac{2}{3}\,\alpha \, \delta ^2 \,(A+B)$, $b_{2}=\frac{4}{3}\, \alpha ^2 \,\delta \,(B-2 A) $ and $b_{3}=2 \,\alpha^{3} \,\Big( \frac{(B-A)^{2}}{A} + (A+B) \Big)$ for NS-NS flux and $a_{0}=2 \,\delta ^3 \,A $, $a_{2}=\frac{2}{3} \,\alpha ^2 \,\delta \,(B-2 A) $ for R-R flux. This family of solutions gives rise to $g_{s}>1$ for $A,B>0$ and then, non-perturbative string effects cannot be neglected.

\section{Summary.}

In this work we have studied the role played by the non-geometric fluxes in type IIB string theory compactified on the $\mathbb{T}^{6}/\mathbb{Z}_{2} \times \mathbb{Z}_{2}$ orientifold with O3/O7-planes. We discussed the role of the non-geometric $Q$ flux which arises from considering T-duality on orbifolds with a non-trivial NS-NS $\bar{H}_{3}$ background flux and the need to match type IIA and IIB effective theories. We then introduced a non-geometric $P$ flux, produced as a S-duality partner of $Q$. Centering our discussion on the isotropic space, so as to deal with only the three moduli fields $T$, $S$ and $U$, we followed the work initiated by \cite{Font:2008vd} and developed methods to address the new scenarios $S$-duality introduced to the space. 
\\

The non-geometric $Q$ and $P$ together with the NS-NS $\bar{H}_{3}$ and R-R $\bar{F}_{3}$ background fluxes were known to define a twelve dimensional algebra, resulting from the reduction of the original metric and $B$ field, with the diffeomorphisms and gauge backgrounds symmetry generators being the generators of the Lie algebra. A complete classification of the gauge Lie subalgebras compatible with the orientifold symmetries was carried out in \cite{Font:2008vd} for the T-duality invariant effective theory, without the $P$ flux. A systematic approach to solving Bianchi identities arising from both the six dimensional gauge subalgebra and the twelve dimensional T-duality invariant algebra (which results in a linear system), was explained there, as well as relations between physically viable vacua with net charges of localized sources as O-planes and D-branes.
\\

At this point we went one step further and investigated the effect of applying S-duality transformations to the constraints on fluxes, ie. Bianchi identities and tadpole cancellations, in the T-duality invariant effective theory. New Bianchi identities appeared involving the non-geometric $Q$ and $P$ fluxes and we were able to identify them as integrability and cohomology conditions needed for the $P$ flux to define deformations of the T-duality invariant gauge subalgebra by an element of its second cohomology class.
\\

The problem of solving the integrability condition forced the non-geometric $P$ flux to define another six dimensional Lie algebra compatible with the orientifold symmetries, in analogy with the non-geometric $Q$ flux. Even though both Lie algebras for $Q$ and $P$ fluxes could be chosen independently, their embeddings were restricted by the cohomology condition. This restriction, far from being trivial to fulfil, determined the form of the modular transformation matrices $\Gamma_{Q}$ and $\Gamma_{P}$, defining how the Lie algebras could be embedded within isotropic fluxes.
\\

At this point, algebraic geometry techniques were required. We made extensive use of the free software \textit{Singular} to compute all solutions to the cohomology condition, breaking it into several families or branches with different physical implications. Different branches of solutions to the integrability and cohomology conditions were interpreted geometrically, as root alignments between the non-geometric flux-induced polynomials entering into the effective superpotential.
\\

New Bianchi identities of the twelve dimensional algebra involving NS-NS $\bar{H}_{3}$ and R-R $\bar{F}_{3}$ background fluxes remained a linear system. This fact allowed us to split non-geometric background fluxes into what we referred to as type A and B configurations. The type B configurations were found to be those for which a non vanishing flux-induced $C'_8$ tadpole might be generated.
\\

To clarify each step along the paper, we always worked with an example, namely $\,\mathfrak{g}_{Q}=\mathfrak{su(2)}+\mathfrak{u(1)^{3}}\,$ for the non-geometric $Q$ flux, with the algebra for the $P$ flux initially being the trivial case of $\mathfrak{u}(1)^{6}$ before moving onto other, richer, examples as our results developed, particularly $\mathfrak{so}(4)$, which illustrated many important physical vacua properties. Using our methods for that algebra pairing, supersymmetric solutions were simple to compute. We presented a simple AdS$_{4}$ solution with all moduli stabilized and for which no localized sources are needed at all to cancel tadpoles, as well as several supersymmetric Minkowski solutions. Because of the importance of the latter from the phenomenological point of view, we further presented some families of supersymmetric type B Minkowski solutions based on different pairings and identified examples already found in the literature within our construction. Supersymmetric Minkowski solutions can be easily and systematically worked out using these methods.
\\

The methods developed in this work can be extended to a general non-isotropic setup or even to a different orbifold once the six dimensional algebras compatible with the orbifold symmetries have been identified. However, less restricted orbifolds would increase the number of fluxes and probably translate into higher computational costs when using \textit{Singular}. Even so, the ever increasing power of computers is making such paths of research more and more feasible.
\\

We consider that this work could have two main extensions:

\begin{enumerate}
\item From the phenomenological point of view, a deeper study of vacua associated to each pairing of non-geometric fluxes would be interesting to work out as well as a relation between deformed algebras, well defined vacua and the necessity of certain types of localized sources. It is also interesting to explore possible non supersymmetric solutions, their vacuum energy and SUSY breaking scale, as well as their mass spectra.

\item From a more theoretical approach, the identification of deformed algebras induced by the non-geometric $P$ flux has not been solved. In general, once the $\mathfrak{g}_{Q}$ algebra is deformed by $\mathfrak{g}_{P}$, the resulting algebra is no longer isomorphic to the original one (unless the deformation becomes trivial). Also extending this approach to the twelve dimensional algebra structure including $\bar{H}_3$ and $\bar{F}_3$ background fluxes could clarify some aspects about the uplifting to ten dimensions.
\end{enumerate}

In this work, we solved the Bianchi identities involving the non-geometric fluxes induced by S-duality, viewing them as linear deformations of the gauge subalgebra of the T-duality invariant effective theory. However, just a small subset of those solutions were found to survive in the $SL(2,\mathbb{Z})^{7}$-duality invariant supergravity (see table \ref{tablegQgP}). Analyzing in depth the origin of such solutions would also be interesting and is postponed for future work.
\\

%---------------------------------------------------------------------------------------------------------------------------------%

\textbf{\begin{large}Acknowledgments\end{large}}
\\

We are grateful to B.~de Carlos and A.~Font for useful comments and references and, in particular, to P. G. C\'amara for useful discussions and sharing with us a preliminary version of \cite{acr} and J.~M.~Moreno for valuable discussions and comments. A.G. acknowledges the financial support of a FPI (MEC) grant reference BES-2005-8412 as well as the School of Physics and Astronomy, University of Southampton, for hospitality and support at several stages of this paper. G. J. W. is grateful for the support of a Scholarship from the University of Southampton and also the Instituto de F\'isica Te\'orica UAM/CSIC, for hospitality and support. This work has been partially supported by CICYT, Spain, under contract FPA 2007-60252, the European Union through the Marie Curie Research and Training Networks "Quest for Unification" (MRTN-CT-2004-503369) and UniverseNet (MRTN-CT-2006-035863) and the Comunidad de Madrid through Proyecto HEPHACOS S-0505/ESP-0346.
\\

\end{document}